%% file: inpainting_revisited.tex
\documentclass[journal]{IEEEtran}
\ifCLASSINFOpdf
   \usepackage[pdftex]{graphicx}
   \graphicspath{{figures/},{figures/old/}}
   \DeclareGraphicsExtensions{.pdf,.jpeg,.png}
\else
   \usepackage[dvips]{graphicx}
   \graphicspath{{figures/}, {figures/old/}}
   \DeclareGraphicsExtensions{.eps}
\fi
\usepackage{amsmath}
\usepackage[algoruled,linesnumbered]{algorithm2e}

\ifCLASSOPTIONcompsoc
\usepackage[caption=false,font=normalsize,labelfont=sf,textfont=sf]{subfig}
\else
\usepackage[caption=false,font=footnotesize]{subfig}
\fi
\usepackage{caption}

\usepackage{stfloats}
\usepackage{url}
\usepackage{hyperref}

\usepackage{amssymb,amsthm}
\usepackage[utf8]{inputenc}
\usepackage[usenames, dvipsnames]{color}
\usepackage{xcolor}

\include{defs}

\begin{document}
%\title{Audio Inpainting Revisited\todo{: All About That Weight\\ 
\title{Audio Inpainting: Revisited and Reweighted}
% a~structure at a~~ so this keeps an~author's name from being broken across
% a~separate \thanks must be used for each paragraph as LaTeX2e's \thanks

\author{Ondřej~Mokrý
		and
        Pavel~Rajmic
        \thanks{O. Mokrý and P. Rajmic are with the SPLab at the Faculty of Electrical Engineering and Communication, Brno University of Technology, Czech Republic. E-mail: \texttt{170583@vutbr.cz}, \texttt{rajmic@feec.vutbr.cz}}
        %and~Zdeněk~Průša\thanks{Z. Průša is with Acoustic Research Institute, Austrian Academy of Sciences, Austria}
    }% <-this % stops a~space
% \thanks{M. Shell was with the Department
% of Electrical and Computer Engineering, Georgia Institute of Technology, Atlanta,
% GA, 30332 USA e-mail: (see http://www.michaelshell.org/contact.html).}% <-this % stops a~space
% \thanks{J. Doe and J. Doe are with Anonymous University.}% <-this % stops a~space
% \thanks{Manuscript received April 19, 2005; revised August 26, 2015.}}

% these prevent an~unwanted space from occurring between the last author name
% a~space would be appended to the last name and could cause every name on that
% that ends a~line with a~% and do not let a~space in before the next \thanks.
% this is a~minor point as most people would not even notice if the said evil

\markboth{IEEE/ACM Trans. on Audio, Speech, and Language Processing}%
{Shell \MakeLowercase{\textit{et al.}}: Bare Demo of IEEEtran.cls for IEEE Journals}

% Puvodni header:


% If you want to put a~publisher's ID mark on the page you can do it like


\maketitle

% As a~general rule, do not put math, special symbols or citations
\begin{abstract}
%\todo{Puvodni abstrakt}
%We deal with the problem of sparsity-based audio inpainting.
%A~consequence of optimization approaches is actually the insufficient energy of the signal within the filled gap.
%We propose improvements to the audio inpainting framework based on sparsity and convex optimization,
%aiming at compensating for this energy loss.
%The new ideas are based on different types of weighting, both in the coefficient and the time domains.
%We show that our propositions improve the inpainting performance both in terms of the SNR and ODG.
%However, the autoregressive Janssen algorithm remains a~strong competitor.
%% First of the proposed methods is weighting the coefficients of Gabor representation of the signal before running the optimization algorithm. Second method is processing the gap gradually, shortening the gap step by step until the whole gap is filled. Third method is time domain processing, where we search for a suitable curve to multiply the filled gap with to compensate for the energy loss. All methods are tested on same set of signals and the results are evaluated using signal-to-noise ratio (SNR). It is shown that generally, the proposed methods reach SNR comparable to the basic algorithm, or even significantly higher for sparse signals.
%%
%%\todo{All is based on weighting, either in the transformed domain or the time domain.}We deal with the problem of sparsity-based audio inpainting.
%\\
%\todo{Novy navrzeny abstrakt}
\edit{We deal with the problem of sparsity-based audio inpainting, i.e.\ filling in the missing segments of audio.
A~consequence of the approaches based on mathematical optimization is the insufficient amplitude of the signal in the filled gaps.
Remaining in the framework based on sparsity and convex optimization, we propose improvements to audio inpainting,
aiming at compensating for such an energy loss.
The new ideas are based on different types of weighting, both in the coefficient and the time domains.
We show that our propositions improve the inpainting performance in terms of both the SNR and ODG.
%However, the autoregressive Janssen algorithm remains a~strong competitor.
% First of the proposed methods is weighting the coefficients of Gabor representation of the signal before running the optimization algorithm. Second method is processing the gap gradually, shortening the gap step by step until the whole gap is filled. Third method is time domain processing, where we search for a suitable curve to multiply the filled gap with to compensate for the energy loss. All methods are tested on same set of signals and the results are evaluated using signal-to-noise ratio (SNR). It is shown that generally, the proposed methods reach SNR comparable to the basic algorithm, or even significantly higher for sparse signals.
%
%\todo{All is based on weighting, either in the transformed domain or the time domain.}
}
\end{abstract}


%\tableofcontents

% \fi
%
% For peerreview papers, this IEEEtran command inserts a~page break and
\IEEEpeerreviewmaketitle

\section{Introduction}

% The very first letter is a~2 line initial drop letter followed
% form to use if the first word consists of a~single letter:
% Here we have the typical use of a~"T" for an~initial drop letter

\IEEEPARstart{A}{udio} inpainting deals with missing samples in digital audio signals.
Different algorithms were developed aiming at the restoration of the lost information.
In practice, the typical loss of signal is in the form of a~compact gap,
for instance due to a~dropout in Voice-over-IP communication.

The methods proposed by Janssen \cite{javevr86,Oudre2018:Janssen.implementation} and Etter \cite{Etter1996:Interpolation_AR} are among the oldest (but most successful!) methods.
They are based on autoregressive signal modeling;
the missing samples are filled by linear prediction using autoregressive coefficients that are learned using the neighborhood of the gap.
For a more comprehensive study into AR-based audio inpainting, see, for example, \cite{Esquef2003:Interpolation.Long.Gaps.Warped.Burgs, kupinnen2002:extrapolation, Kauppinen2002:reconstruction.method.long.portions.audio}.
\edit{Approaches based on statistical methods were presented in \cite{Godsill1998:Digita.Autio.Restoration} for the related problem of click removal,
and the Bayesian approach to inpainting/declipping was introduced in \cite{Chantas2018:Inpainting_Variational_Bayesian_Inference}.}

A range of audio processing applications came along with the advent of sparse signal representations \cite{eladbook,DonohoElad2003:Optimally}.
The first work that used sparse signal synthesis for filling the missing samples \cite{Adler2012:Audio.inpainting}
actually took over the term \qm{audio inpainting} from the image processing field.

For gaps longer than approximately 100 milliseconds, all the above-described approaches start to fail.
The main reason is that audio can usually be considered stationary only for a few tens of milliseconds.
That is why other modeling approaches have been introduced for longer gaps:
(generalized) sinusoidal modeling \cite{lagrange2005long,Lindblom2002:PLC.sinusoidal}, similarity graph approach
\cite{Bahat_2015:Self.content.based.audio.inpaint,Perraudin2018:Similarity.Graphs}
or deep neural networks
\cite{Marafioti2019:Context.encoder,Marafioti2019:DNN.inpainting}.
%http://www.aes.org/e-lib/browse.cfm?elib=20303
%https://ieeexplore.ieee.org/document/8867915

In the present article, we concentrate on the classic case where the gap does not exceed 50 milliseconds.
We emphasize that we assume filling a~single, compact gap.
%Filling a~compact gap is a~real
Such a task is truly challenging, which explains why not many methods have been published on the topic,
at least in comparison with the related field of audio declipping.
Clipping is a non-linear distortion that degrades the signal all along its length;
there is no compact-in-time loss, and moreover, some information is still available due to the knowledge of the clipping model.
For the treatment of clipping, much more literature exists
\cite{Adler2011:Declipping, Kitic2015:Sparsity.cosparsity.declipping, BilenOzerovPerez2015:declipping.via.NMF, RenckerBachWangPlumbley2018:Consistent.dictionary.learning.LVA, ZaviskaRajmicPrusaVesely2018:RevisitingSSPADE, ZaviskaRajmicMokryPrusa2019:SSPADE_ICASSP}, to name but a~few.
Note that another related problem is filling in the missing samples which are selected randomly, as done,
for instance, in \cite{Lieb2018:Audio.Inpainting}.
In such a case, the situation is close to clipping; inpainting is then relatively simple, since the occurrence of a~significantly long sequence of missing samples is highly improbable.

For our scenario, where we treat short gaps, successful methods are typically model-based;
an optimization problem is designed that contains the data fitting term and a regularizer.
The regularizer usually penalizes the deviation of transformed signal's coefficients from the model under consideration.
The transform used here is typically a~kind of short-time spectral transform.

\begin{figure}[h]
	\centering
	\includegraphics[width=0.9\linewidth]{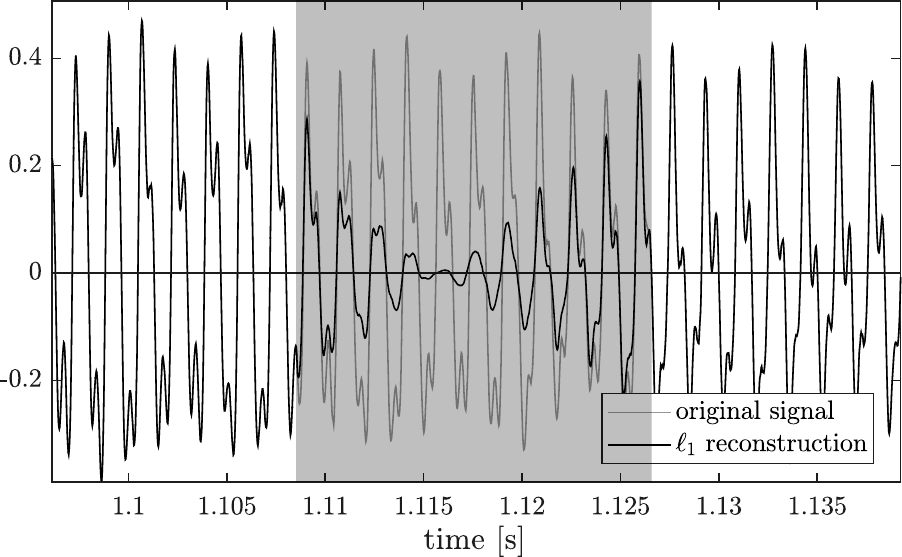}
	\caption{Illustration of a~typical energy loss inside the filled gap. The gap is visualized as a gray area throughout the paper.}
	\label{fig:energy_decrease}
\end{figure}

A problem with these approaches is that regularizers make the solution biased.
This effect will be discussed later on, but let us reveal right now that in the case of audio inpainting, the bias manifests itself in the form of a signal's energy drop within the filled gap, see Fig.\,\ref{fig:energy_decrease}.
The main goal of the article is to study this effect,
design a number of methods that all aim at compensating it,
and evaluate them in numerical experiments on a real audio.

%As usual in audio, signals will be processed window by window.
In Sec.\,\ref{Sec:Gabor.systems}, we summarize the basics of Gabor transforms that will be used for window-wise signal processing.
Sec.\,\ref{sec:synthesis_analysis} then inspects the synthesis and analysis models of sparse audio inpainting
and introduces different types of weighting the signal coefficients.
Sec.\,\ref{sec:reweighting} follows up with the idea that the weights are iteratively recomputed.
Sec.\,\ref{sec:offset} presents the unexpected effect of the shift of the Gabor system in time on the result of inpainting.
A gradual approach is described in Sec.\,\ref{sec:gradual}, where the gap is not filled at once but piece by piece, inspired by image inpainting methods.
Sec.\,\ref{sec:tdc} then closes the list of proposed algorithms with the time-domain approach.
Finally, experiments are conducted and evaluated in Sec.\,\ref{sec:exp}.

Note that we are not aware of any paper that would study the weighting of signal coefficients, apart from our preliminary study \cite{MokryRajmic2019:Reweighted.l1.inpainting}. %\todo{a normování ufiknutých atomů v \cite{Adler2012:Audio.inpainting}?}
Maybe even more surprisingly, there is no audio inpainting article that would involve the \emph{analysis} model, except the evaluation study \cite{Lieb2018:Audio.Inpainting}.

\section{Gabor systems}
\label{Sec:Gabor.systems}
%\dod{tato část je adept na (drastické) zkrácení}\\
%\dod{ale je třeba zachovat definici pojmu atom}
The audio inpainting model presented hereinafter is based on the time-frequency sparsity of audio signals.
As the sparsifying transform, we use the Gabor transform, known also as the Short-time Fourier transform (STFT).
%we need to use a transform that will provide us a sparse representation of the audio signal.
%We will make use of the observation that audio signals are mainly composed of relatively small number of harmonic frequencies.
The Gabor transform uses a~single prototype window function which is translated in time and frequency-modulated
\cite{Grochenig2001:Foundations.T-F.analysis,christensen2008}.
The window serves to localize the signal in time.
In discrete time, the translation of the window, denoted $\g$, is characterized by the integer window shift $a$.
The number of modulations
%per each translation of window 
is denoted $M$ and will be referred to as the number of frequency channels.

There exist convenient combinations of $\g$ and the parameters $a,M$ such that the resulting Gabor system forms a~frame of $\CC^L$, i.e.\
%there exist constants $0<A\leq B<\infty$ for which inequality
%%
%\begin{equation}
	%A\norm{\y}_2^2 \leq \norm{\ana\y}_2^2 \leq B\norm{\y}_2^2
	%\label{eq:frame}
%\end{equation}
%%
%\todo{G s hvezdou nebylo zavedeno}
%holds for all $\y \in \CC^N$
%
any \edit{signal} $\y\in\CC^L$ is representable in a stable way as the linear combination of the Gabor vectors %of the system
\cite{christensen2003,heil,Feichtinger_gabor2001}.
Although Gabor bases can be constructed,
% they have undesirable properties and 
overcomplete systems which allow non-unique signal representation are preferred.
In the overcomplete setup, the synthesis operator $\syn\colon\CC^N\rightarrow\CC^L$ generates a signal \edit{of length $L$ from the $N>L$} coefficients.
Its adjoint, \edit{the analysis operator} $\ana\colon\CC^L\rightarrow\CC^N$, produces coefficients out of the signal.
%In discrete time, $\syn$ and $\ana$ can be regarded matrices, and application of these operators amounts to taking inner products.
Note that we treat the signal as a~complex vector,
although we work only with real signals.
%but since we work only with real signals, we suppose that its imaginary part is zero.
% As a consequence, the analysis of a real signal produces coefficients in conjugate pairs, in order to produce real signal by applying the synthesis operator $\syn$.

%In this paper, exclusively Gabor frames for which $A = B = 1$ will be used.
In this paper, only Gabor \emph{tight frames} will be used.
They have convenient properties from both the theoretical and practical points of view.
%and at they are at the same time they are \todo{sufficient} in applications.
For example, the windows used in synthesis and analysis are identical (up to a~scale factor).
We will make use of the so-called Parseval tight frames, which
in addition
% are easily obtained by scaling to 
satisfy \cite{christensen2008}% \todo{ocitovat nebo vysvětlit}
\begin{equation}
	\syn\ana = \Id,\quad
	\norm{\ana\y}_2 = \norm{\y}_2,
	\label{eq:tight.frame.analysis.retains.norm}
\end{equation}
%
%in other words,
%i.e.\
%the analysis operator retains the signal norm.
%
\edit{where $\Id$ in general denotes the identity operator; in this particular context, it is the identity on the space $\CC^L$.}

%that form tight frames, i.e.\ for which there exist constants $A = B$ for which \eqref{eq:frame} holds, and especially Parseval tight frames with $A = B = 1$ \todo{[citation: něco o framech]}.
%\todo{Dulezity dusledek je, ze syn. okno je jen nasobkem analyzujiciho, pro Parsevala dokonce identicke.
	%Mj. to tedy znamena, ze maji stejnou delku.}

%Since we will be searching for sparsest representation under certain conditions, we need the system used to represent the audio signal to be overcomplete.
%These requirements are satisfied by time-frequency representations termed Gabor systems, known as the Short-time Fourier transform (STFT) as well.

%\todo{Sice existují báze, ale jsou na nic, a hlavně potřebujeme syn a analýzu}

Going into greater detail,
it is natural to require that the signal length $L$ should be divisible by $a$.
In \edit{this particular case, the Gabor system consists of $N = \frac{L}{a}\cdot M$} vectors $\{\vect{d}_n\}_{n=1,\dots,N}$,
$\vect{d}_n\in\CC^L$.
We will refer to these vectors as the (frame) \emph{atoms} and to the whole system as the \emph{Gabor dictionary}.

The window $\g$ is usually identified with its shorter counterpart,
built by keeping only the nonzero samples of $\g$.
%, whose number is usually much lower than $L$.
%We denote the effective support with respect to signal length $p$.
Therefore we refer to the support length $w := \abs{\supp(\g)}$ \edit{(i.e.\ the number of non-zero elements of $\g$)} as the \emph{window length}.
%\edit{By $|\supp(\x)|$, we denote the cardinality of the support of $\x$, i.e., the number of non-zero elements of $\x$.}
We will work exclusively with finitely-supported windows in this article.

We use the fast implementation of Gabor transforms offered by the LTFAT toolbox
\cite{LTFAT,ltfat-web} in our computations, and we adopt its time-frequency conventions.

\section{Synthesis and analysis models with weights}
%\dod{zvazit jestli nazev je dobry}\\
%\dod{mozna misto A pouzivat pouze D s hvezdickou}
\label{sec:synthesis_analysis}
The sparse signal processing literature
%, including the audio processing branch,
relied for long
%about two decades ago with
on the so-called synthesis model, where one seeks for a~small number of coefficients which are then synthesized to produce the resultant signal
\cite{DonohoElad2003:Optimally,chen,Tibshirani1996:Regression,Bruckstein.etc.2009.SIAMReviewArticle,AharonEladBruckstein2006:KSVD}.
More recently, the analysis model has been studied, where one looks directly for the signal, with the requirement that its coefficients after analysis should be sparse
\cite{Kitic2015:Sparsity.cosparsity.declipping,Elad05analysisversus}.
Both approaches are equivalent if the synthesis/analysis operators are bijective,
i.e.\ if the operators correspond to the bases for the signal/coefficient spaces.

In this section we introduce these two approaches in the audio inpainting problem.
Besides, we explore several methods for atom weighting, in order to improve the performance of the restoration.
%We show that the difference among the models under consideration starts to reveal especially when they are applied to signals with denser TF representation.
%We show that there is not a significant difference if we compare between the results of synthesis and analysis model when we are restoring sparse signal.
%On the contrary, if the signal is less sparse, we show that the difference becomes significant and it is dependent on the weighting method.

\subsection{Problem formulation}

\edit{%
Let $\y$ denote
the time-domain signal.
%and $\x$ is the vector of Gabor coefficients.
Let the indexes of missing (or unreliable) samples be known.
This will be referred to as \emph{the gap}.
The rest of the samples will be considered non-degraded and will be called \emph{reliable}.
}

\edit{%
It is natural to require that the recovered signal maintains consistency with the observed signal in the reliable part.
%In this context,
To formally handle this requirement,
we introduce the (convex) set $\Gamma$, the set of all feasible signals
%satisfying the aforementioned constraint, formally
%
\begin{equation}
\Gamma = \lbrace\z\in\CC^L\mid\Mr\z=\Mr\y\rbrace,
\end{equation}
%
%\todo{signals real}
where $\Mr\colon\CC^L\to\CC^L$ is the \qm{reliable mask} projection operator.
	It maps a signal in $\CC^L$ to another signal in $\CC^L$,
keeping the signal samples corresponding to the reliable part intact,
while setting the others to zero.
}

The sparse audio inpainting can be formulated as a~minimization problem.
In the case of the linear synthesis model, we assume that the synthesis operator $\syn$
%is overcomplete and has full rank, which means
allows any signal from $\CC^L$ to be generated from an infinite number of choices of coefficients
(due to the overcompleteness of the system).
%(cf.\ definition of the frame in Sec.\,\ref{Sec:Gabor.systems}).
The sparse synthesis model aims at obtaining the highest sparsity representation that fits the reliable signal samples, formally
%would are searching for a~signal with the sparsest Gabor representation in the set $\Gamma$ of signals equal to the one we are restoring in its reliable part (i.e.\ except the missing samples). The problem can be written as
%
\begin{equation}
	\label{eq:inpainting_orig}
	\argmin_\x \norm{\x}_0 \quad \text{s.t.}\ \syn\x\in\Gamma ,
\end{equation}
where
%$\syn:\CC^N\rightarrow\CC^q$ denotes the Gabor synthesis operator, % throughout the paper.
%and 
$\norm{\cdot}_0$ denotes the $\ell_0$-pseudonorm, which simply counts the non-zero elements of the argument.
\edit{Note that this is the same quantity as $|\supp(\cdot)|$.
However, we keep both notations throughout the paper, since $|\supp(\cdot)|$ emphasizes the measurment of length in the time domain.}

Solving optimization problems that involve $\norm{\cdot}_0$ is NP-hard and thus computationally intractable.
%since sparsity is not a~convex function,
Therefore approximations of the true solution to \eqref{eq:inpainting_orig} must be introduced.
Probably the most common way today is to solve a~relaxed minimization problem
that involves the $\ell_1$ norm instead of $\ell_0$ \cite{DonohoElad2003:Optimally,FornasierEditor2010:SparseRecoveryBook},
allowing the use of convex optimization
\cite{chen,Boyd:2004:Book:ConvexOptimization}.

In the formulations that follow we include the weighting vector $\w\in\RR^N,\,\w>0$.
Its role is to assign potentially different weights to the coefficients, leading to a minimization of the
%so-called
weighted $\ell_1$ norm. %\todo{\textit{zavedeme speciální značení? budeme psát spojovník?}}.
The relaxed synthesis formulation reads
\begin{equation}
	\argmin_\x \norm{\w\odot\x}_1 \quad \text{s.t.}\ \syn\x\in\Gamma,
	\label{eq:inpainting_syn_constrained}
\end{equation}
where $\norm{\cdot}_1$ sums the magnitudes of elements of the argument, and $\odot$ denotes the entrywise (Hadamard) product.
Problem \eqref{eq:inpainting_syn_constrained} can be equivalently written in an~unconstrained form as
\begin{equation}
	\argmin_\x \norm{\w\odot\x}_1 + \iota_\Gamma(\syn\x),
	\label{eq:inpainting_syn_unconstrained}
	\tag{AIs}
\end{equation}
where $\iota_C$ is the indicator function of a~set $C$---it takes on zero value for elements belonging to $C$ and infinity otherwise.

On the other hand, %contrary,
the analysis formulation is
\begin{equation}
\argmin_\z \norm{\w\odot\ana\z}_1 \quad \text{s.t.}\ \z\in\Gamma,
\label{eq:inpainting_ana_constrained}
\end{equation}
or, in an~unconstrained form,
\begin{equation}
	\argmin_\z \norm{\w\odot\ana\z}_1 + \iota_\Gamma(\z).
	\tag{AIa}
	\label{eq:inpainting_ana_unconstrained}
\end{equation}
It is clear that the output of the analysis minimization
\eqref{eq:inpainting_ana_unconstrained}
is directly the restored signal.
On the contrary, the synthesis model
\eqref{eq:inpainting_syn_unconstrained}
finds the optimal vector of coefficients, and the restored signal is obtained simply via the application of $\syn$.
 %to it.

\subsection{Choosing the weights}
\label{sec:atom_weighting}

A consequence of the $\ell_1$ minimization is that not only a~number of coefficients are pushed to zero, in order to make the coefficient vector sparse, but also the non-zero coefficients
%important for the reconstruction
are automatically made smaller in magnitude than they could (and should) be.
This is a~commonly observed problem, which is called \emph{bias} in the statistical community
\cite{HastieTibsWainw2015:Statistical.learn.sparsity,CandesWakinBoyd2008:enhanced.reweighted.l1,Rajmic2003:Exact.risk.analysis,Dankova2016}.

\edit{
In the context of the Gabor transform with a system of translated windows, notice that besides the just described global effect of the $\ell_1$ norm,
the coefficients corresponding to windows that overlap with the gap carry less information about the reliable signal than the coefficients corresponding to the reliable parts.
 %and thus are more affected by the bias.
%The larger the window overlap with the gap is, the smaller the coefficients are. %since the coefficients are basically only the scalar products.
As a~consequence,
%a~gradual
a~progressive
loss of amplitude is typically observed in the restored signal,
as already demonstrated in Fig.\,\ref{fig:energy_decrease}.
}

%\begin{figure}[h]
%	\centering
%	\includegraphics[width=0.85\linewidth]{energy_loss}
%	%
%	\caption{Illustration of a~typical energy loss inside the filled gap. The gap is visualized as the gray area throughout the paper.}
%	%
%	\label{fig:energy_decrease}
%\end{figure}

In order to compensate for this local effect,
we naturally propose weighting the frame atoms---the less reliable information the atom carries,
the lower the corresponding weight should be, resulting in less penalization in either
\eqref{eq:inpainting_syn_unconstrained}
or
\eqref{eq:inpainting_ana_unconstrained}.
\edit{%
	Note that such an idea already appeared in \cite{Adler2012:Audio.inpainting}, where the authors used the weighting such that the $\ell_2$ norms of atoms used for inpainting were normalized.
	In this article, we propose and examine several other ways to determine the weighting coefficients.
	For the particular choices, see Sec.\,\ref{sec:exp_weighting}.
}

%By setting the weights this way, the atoms important for inpainting the gap will not be harmed so much by minimization of the $\ell_1$ norm in problems
%\eqref{eq:inpainting_syn_unconstrained_gamma_ad} and \eqref{eq:inpainting_ana_unconstrained}.

\subsection{Proximal algorithms}
\label{Sec:Proximal}
The proximal splitting methodology is an~efficient tool for iterative solution to large-scale convex minimization problems
\cite{combettes2011proximal,ChambollePock2011:First-Order.Primal-Dual.Algorithm,KomodakisPesquet2015:Playing.with.duality,Condat2014:Generic.proximal.algorithm}.
Certain proximal algorithms are able to find the minimum of a~sum of convex functions $f_i$, with mild assumptions about these functions, even when some of the functions $f_i$ are composed with linear operators.
%Based on the properties of the functions,
Proximal algorithms perform iterations involving an evaluation of the so-called \textit{proximal operators} related individually to each $f_i$, which is computationally much simpler than minimizing the composite functional by other means.
%Proximal algorithms provide sequences that are guaranteed to converge to the optimal value.
%The speed of convergence is influenced by the properties of particular $f_i$s and by particular values of the parameters chosen in the algorithms.
%In the audio inpainting context, we will exploit two particular simple proximal algorithms which are explained below.
%But first of all, it will be convenient to introduce the building blocks that will be used in the algorithms.
%
We will use proximal algorithms to numerically solve our problems
\eqref{eq:inpainting_syn_unconstrained}
and
\eqref{eq:inpainting_ana_unconstrained}.

The proximal operator of a~(convex) function $h\colon\CC^N\rightarrow\RR$ is the mapping
%$\z\in\CC^N$ to another vector in $\CC^N$,
$\prox_h \colon \CC^N \rightarrow \CC^N$ \cite{combettes2011proximal,Moreau1965:Proximite.dualite}.
%such that
%%
%\begin{equation}
	%\label{Eq:Proximal.operator.definition}
	%\prox_h (\z) = \argmin_{\x\in\CC^N} \ \left\{  h(\x) + \frac{1}{2}\norm{\z-\x}_2^2 \, \right\} .
%\end{equation}
%\dod{to by asi slo vypustit, ale ponechat znaceni prox}
%
This article will make use of the proximal operators of two particular functions.
The proximal operator corresponding to the $\ell_1$ norm is the well-known soft thresholding \cite{Donoho1995_De-noising.by.soft.thresholding}.
%When $h=\tau\norm{\cdot}_1$, operator $\prox_{\tau\norm{\cdot}_1}$ is 
%the well-known soft thresholding \cite{Donoho1995_De-noising.by.soft.thresholding},
%denoted $\textstyle{\soft_\tau}$ and defined as an elementwise mapping
%%
%\begin{equation}
	%\label{eq:Soft.thresholing}
	%%z_{i} \mapsto \sgn(z_{i}) \cdot \max(\abs{z_{i}}-\tau,0)
	%\z \mapsto \arg(\z) \odot \max(\abs{\z}-\tau,0).
%\end{equation}
%%
%%where $\odot$ denotes elementwise multiplication.
%%
%Above we justified the use of the weighting operator inside the $\ell_1$ norm.
Conveniently, if the $\ell_1$ norm is composed with elementwise weighting, $\prox_{\tau\norm{\w\odot\cdot}_1}=: \soft_{\tau\vect{w}}$ can be shown to be another elementwise mapping,
where each $w_n$ only affects the threshold value, such that we write
\begin{equation}
	\soft_{\tau\vect{w}}(\x) := \arg(\x)\odot\max(\abs{\x}-\tau\vect{w},0).
	\label{eq:soft}
\end{equation}
%
%For real inputs, the definitions of soft thresholding involve $\sgn(\cdot)$ as the replacement of the more general $\arg(\cdot)$.
%
\edit{%
The operator $\arg(x)$ denotes the argument of $x$, where $x$ is a~complex number.
In \eqref{eq:soft}, this operation is extended elementwise to vectors.
}
When $h=\iota_C$ is the indicator function of a~convex set $C$, 
the related proximal operator $\prox_{\iota_C}(\x)$ finds the vector in $C$ closest to $\x$.
Such an operator thus corresponds to the projection onto $C$ and will be denoted $\proj_C$.

The last useful property is related to the Fenchel--Rockafellar conjugate \cite{Boyd:2004:Book:ConvexOptimization}.
Given a~convex $f$, the proximal operator of its conjugate $f^*$ can be computed at virtually the same cost as $\prox_{f}$ due to the Moreau identity
\cite{KomodakisPesquet2015:Playing.with.duality,Condat2014:Generic.proximal.algorithm}:
\begin{equation}
	\label{eq:Prox.of.conjugate}
	\prox_{\alpha f^*}(\u) = \u - \alpha\, \prox_{f/\alpha} (\u/\alpha) \quad \text{for\ } \alpha\in\RR^+.
\end{equation}

\subsection{Solving the synthesis problem}
\label{Sec:Solving.synthesis.problem}
%\todo{odvolat se na \eqref{eq:inpainting_syn_unconstrained_gamma_ad}}
%A~general form of algorithm Douglas-Rachford (DR) with a~parameter $\tau$ can be found in \cite{combettes2011proximal}.
Returning to the synthesis-based inpainting \eqref{eq:inpainting_syn_unconstrained},
notice that this problem minimizes the sum of two convex functions
\begin{equation}
	f_1 = \iota_{\Gamma}\circ\syn,\quad f_2 = \norm{\w\odot\cdot}_1.
\end{equation}
It is convenient to use the Douglas--Rachford (DR) algorithm \cite{combettes2011proximal} to find the numerical solution.
The building blocks of the DR algorithm are $\prox_{f_1}$ and $\prox_{f_2}$.
In light of Sec.\,\ref{Sec:Proximal},
these operators are the projection and the generalized soft thresholding, respectively.
In our case, $\prox_{f_1}$ has an explicit form 
%The projection, as a proximal operator of composition of linear operator $D$, can be written as
%
\begin{equation}
	\prox_{\tau f_1} (\x) = \x-\ana\Mr\syn\x+\ana\Mr\y,
	\label{eq:proj}
\end{equation}
due $\syn$ being assumed to be a~tight Parseval frame \cite{combettes2011proximal}.
%
%The following algorithm summarizes the \textbf{DR algorithm for inpainting}.
%
%
%
%\begin{algorithmic}[1]
%	\REQUIRE
%	tight synthesis operator $\syn\colon\CC^N \rightarrow \CC^L$,
%	analysis operator $\ana\colon\CC^L \rightarrow \CC^N$,
%	observed signal $\y$,
%	mask $\Mr$,
%	weighting vector $\w$
%	\STATE choose parameter $\tau > 0$
%	\STATE choose auxiliary variable $\q^{(0)}\in\CC^N$ arbitrarily
%	\STATE set iteration counter $i = 0$
%	\REPEAT
%		\STATE $\x^{(i)} = \soft_{\tau\vect{w}} (\q^{(i)})$
%		\STATE $\q^{(i+1)} = \x^{(i)} + \ana\Mr\syn(2\x^{(i)}-\q^{(i)}) + \ana\Mr\y$
%		\STATE $i \leftarrow i + 1$
%	\UNTIL{stopping criterion met}
%	\RETURN $\proj_{\Gamma}(\syn\x^{(i)})$
%\end{algorithmic}
%

\begin{algorithm}[b]
	\DontPrintSemicolon
	\SetKwInput{KwRequire}{require}
	\SetKw{KwReturn}{return}	
	\KwRequire{tight synthesis operator $\syn\colon\CC^N \rightarrow \CC^L$,
		%analysis operator $\ana\colon\CC^L \rightarrow \CC^N$,
		observed signal $\y$,
		mask $\Mr$,
		weights $\w$}
	choose parameter $\tau > 0$\\
	choose auxiliary variable $\q^{(0)}\in\CC^N$ arbitrarily\\
	set iteration counter $i = 0$\\
	\Repeat{stopping criterion met}
	{
		$\x^{(i)} = \soft_{\tau\vect{w}} (\q^{(i)})$\\
		$\q^{(i+1)} = \x^{(i)} + \ana\Mr\syn(2\x^{(i)}-\q^{(i)}) + \ana\Mr\y$\\
		$i\leftarrow i+1$\\
	}
	\KwReturn{$\proj_{\Gamma}(\syn\x^{(i)})$}
	\caption{DR algorithm for inpainting}
	\label{alg:douglasrachford}
\end{algorithm}

The DR algorithm for inpainting is summarized by Alg.\,\ref{alg:douglasrachford}.
The algorithm converges for any positive $\tau$,
but this parameter can largely affect the convergence speed.
In the experiments that will follow, we use
% fixed value of
%$\tau = \frac{1}{10}$
%$\tau = 0.1$
%and
the usual termination criterion
$\norm{\x^{(i)}-\x^{(i-1)}}<\varepsilon\norm{\x^{(i-1)}}$,
where $\varepsilon > 0$ is a~chosen tolerance.
%We also fixed the parameter $\tau = \frac{1}{10}$, since it affects the convergence speed, which is out of interest of this paper.

\subsection{Solving the analysis problem}
\label{Sec:Solving.analysis.problem}
For solving \eqref{eq:inpainting_ana_unconstrained},
%the DR algorithm can no longer be used.
%We shift to
we use the Chambolle--Pock (CP) algorithm \cite{ChambollePock2011:First-Order.Primal-Dual.Algorithm}
%While solving the inpainting problem \eqref{eq:inpainting_ana_unconstrained},
with the assignment
\begin{equation}
	f_1 = \iota_{\Gamma},\quad f_2 = \norm{\w\odot\cdot}_1.
\end{equation}
%
%Notice that
The difference from the synthesis variant is that the argument of $f_2$ is the vector of coefficients of the signal after analysis by $\ana$.
There is no explicit formula like \eqref{eq:proj} in this case and therefore 
%That is the reason why
the DR algorithm is not applicable here.
%the CP algorithm allows the composition of a~linear operator and a~convex functional.

It is of advantage that
% Conveniently, the proximal operator of $f_1$ is the projection
$\prox_{f_1}(\cdot) = \proj_{\Gamma}(\cdot)$,
% being an elementwise operation that replaces the current samples at reliable positions
% by the corresponding samples of the observed signal while preserving the samples in the gap.
since this is an elementwise operation.
Specifically, the current samples in reliable positions are replaced by the corresponding samples of the observed signal while the samples in the gap are preserved.
Formally,
$
	\proj_{\Gamma}(\x) = (\Id-\Mr)\x + \Mr\y.
$
%
%Furthermore, for derivation of the CP algorithm for inpainting, we need the proximal operator of a~Fenchell-Rockafellar conjugate $f_2^*$ of function $f_2$. This can be computed by Moreau identity as
%\begin{equation}
	%\prox_{\sigma h^*}(\x) = \x - \sigma\, \prox_{h/\sigma}(\x/\sigma) \text{ for } \sigma \in \RR^+. 
	%\label{eq:moreau_identity}
%\end{equation}
By defining
%
%\begin{equation}
	$\clip_{\vect{w}}(\x) := \x - \soft_{\vect{w}}(\x)$
	%\label{eq:clip}
%\end{equation}
%
% and using Lemmas \ref{lemma1} and \ref{lemma2} from Appendix~\ref{sec:lemmas},
and using the property
$\soft_{\tau}(\x\cdot\tau) = \tau\cdot\soft_{1}(\x)$ for any $\tau>0$,
we can rewrite Eq.\,\eqref{eq:Prox.of.conjugate} for our $f_2=\norm{\w\odot\cdot}_1$ as
%
%\begin{equation}
%\begin{split}
	%\prox_{\sigma f_2^*}(\x) &=\x - \sigma\cdot \soft_{\vect{w}/\sigma}(\x/\sigma)\\&= \x - \soft_{\vect{w}}(\x) = \clip_{\vect{w}}(\x).
%\end{split}	
%\end{equation}
\begin{equation}
	\prox_{\sigma f_2^*}(\x) = \x - \sigma\cdot \soft_{\vect{w}/\sigma}(\x/\sigma) = \clip_{\vect{w}}(\x).
\end{equation}
%
%The following algorithm summarizes the \textbf{CP algorithm for inpainting}.
%\begin{algorithmic}[1]
%	\REQUIRE
%	tight synthesis operator $\syn\colon\CC^N \rightarrow \CC^L$,
%	analysis operator $\ana\colon\CC^L \rightarrow \CC^N$, \dod{prehodit? nebo vyhodit}
%	observed signal $\y$,
%	mask $\Mr$,
%	weighting vector $\w$
%	\STATE choose $\tau, \sigma>0$ satisfying $\tau\sigma\norm{\ana}^2<1$
%	\STATE choose primal variable $\p^{(0)}\in\CC^L$ and dual variable $\q^{(0)}\in\CC^N$ arbitrarily
%	\STATE set output variable $\y^{(0)}=\p^{(0)}$
%	\STATE set iteration counter $i = 0$
%	\REPEAT
%		\STATE $\q^{(i+1)} = \clip_{\vect{w}}(\q^{(i)}+\sigma \ana \y^{(i)})$
%		\STATE $\p^{(i+1)} = \proj_{\Gamma}(\p^{(i)}-\tau \syn {\q}^{(i+1)})$
%		\STATE $\y^{(i+1)} = 2\p^{(i+1)} - \p^{(i)}$
%		\STATE $i\leftarrow i+1$
%	\UNTIL{stopping criterion met}
%	\RETURN $\proj_{\Gamma}(\y^{(i)})$
%\end{algorithmic}
%
The CP algorithm for inpainting is summarized in Alg.\,\ref{alg:chambollepock}.
\edit{%Note that the condition on the parameters $\tau$ and $\sigma$ is slightly relaxed according to \cite{Condat2014:Generic.proximal.algorithm}.
	%Note that the condition $\tau\sigma\norm{\syn}^2<1$ for its convergence \cite{ChambollePock2011:First-Order.Primal-Dual.Algorithm} is extended according to \cite{Condat2014:Generic.proximal.algorithm}, such that also $\tau\sigma\norm{\syn}^2=1$ is allowed.
%\todo{nerozumím \textbf{Teď už jo? :)}}
The convergence is guaranteed if the step sizes $\tau$ and $\sigma$ are set such that $\tau\sigma\norm{\syn}^2\leq 1$ \cite{Condat2014:Generic.proximal.algorithm}.
}
\vspace{-0.5em}
\begin{algorithm}[h]
	\DontPrintSemicolon
	\SetKwInput{KwRequire}{require}
	\SetKw{KwReturn}{return}	
	\KwRequire{tight synthesis operator $\syn\colon\CC^N \rightarrow \CC^L$,
		%analysis operator $\ana\colon\CC^L \rightarrow \CC^N$, \dod{prehodit? nebo vyhodit}
		observed signal $\y$,
		mask $\Mr$,
		weights $\w$}
	choose $\tau, \sigma>0$ satisfying $\tau\sigma\norm{\syn}^2\edit{\leq}1$\\ %$\tau\sigma\norm{\ana}^2<1$\\
	choose primal variable $\p^{(0)}\in\CC^L$ and dual variable $\q^{(0)}\in\CC^N$ arbitrarily\\
	set output variable $\y^{(0)}=\p^{(0)}$\\
	set iteration counter $i = 0$\\
	\Repeat{stopping criterion met}
	{
		$\q^{(i+1)} = \clip_{\vect{w}}(\q^{(i)}+\sigma \ana \y^{(i)})$\\
		$\p^{(i+1)} = \proj_{\Gamma}(\p^{(i)}-\tau \syn {\q}^{(i+1)})$\\
		$\y^{(i+1)} = 2\p^{(i+1)} - \p^{(i)}$\\
		$i\leftarrow i+1$\\
	}
	\KwReturn{$\proj_{\Gamma}(\y^{(i)})$}
	\caption{CP algorithm for inpainting}
	\label{alg:chambollepock}
\end{algorithm}
\vspace{-0.5em}
We would like to use the termination criterion analogously to the synthesis case,
and therefore we measure the relative difference of the norms
%of analyzed signal, resulting into criterion
$\norm{\ana\y^{(i)}-\ana\y^{(i-1)}}<\varepsilon\norm{\ana\y^{(i-1)}}$,
where $\varepsilon > 0$ is the tolerance.
Notice, however, that the operator $\ana$ is linear and, being a~Parseval tight frame,
it preserves the vector norms (see Eq.\,\eqref{eq:tight.frame.analysis.retains.norm}), hence 
%$\norm{\ana} = 1$ and 
the criterion is equivalent to
$\norm{\y^{(i)}-\y^{(i-1)}}<\varepsilon\norm{\y^{(i-1)}}$.
%\todo{doresit pismenka!}
%
%In our experiments, we set $\tau = \sigma = 1$.
%As in the synthesis model, we are not measuring the convergence speed of the algorithm, therefore we set the parameters $\tau = \sigma = 1$.
%Such choice is possible since  for Parseval tight frame and it has been shown in \cite{Condat2014:Generic.proximal.algorithm} that the algorithm converges also for $\tau\sigma\norm{\ana}^2=1$.

\subsection{Computational complexity}
It is clear from the pseudocodes that both the proximal algorithms perform one analysis (operator $\ana$)
and one synthesis (operator $\syn$) in each iteration---note that the analysis of the reliable part
of the input signal $\y$ also appears in the DR algorithm, but this can be precomputed.
Due to the fact that the complexity of $\syn$ and $\ana$ significantly exceeds the cost of other operations involved,
%From equation \eqref{eq:clip} it is clear that the functions $\clip_{\vect{w}}$ (CP) and $\soft_{\tau\vect{w}}$ (DR) are equivalent regarding the computational cost.
%Because the described algorithms do not contain any other computationally demanding steps, they are equivalent in this manner. 
we can summarize that the CP and the DR algorithms for audio inpainting are identically demanding.

\section{Iterative reweighting}
\label{sec:reweighting}

%\edit{The ongoing sections present the individual ideas for the treatment of the loss of signal amplitude, as presented above.}

In \cite{Weinstein2011:DeclippingSparseland},
audio declipping using the so-called \emph{re}\/weighted $\ell_1$ minimization was presented.
In such an approach, the $\ell_1$ norm of the coefficients is weighted by $\w$ as in the previous section,
%to provide more energy to the restored gap,
but the idea behind the weights is different here:
The restoration task is solved \emph{repeatedly}, and the weights change in the repetitions,
based on the
%inverted absolute
values of the coefficients from the current solution.
The benefit is that using such a~procedure, the significant %significant
coefficients can be adaptively penalized less and less,
while the insignificant %tiny
coefficients are more and more pushed towards zero,
leading to a~better approximation of sparsity (and to avoiding the bias, to some extent).
Note also that simple examples can be found where this strategy fails to find the optimal sparse solution
\cite{CandesWakinBoyd2008:enhanced.reweighted.l1}.

As a matter of fact, %Actually,
\cite{Weinstein2011:DeclippingSparseland} applied the reweighting strategy
only in the synthesis variant of declipping.
We adapted their approach to audio inpainting already in \cite{MokryRajmic2019:Reweighted.l1.inpainting} and include it here for the context.
Recall that the shift of task from declipping to inpainting is done easily
by redefining the set of feasible solutions $\Gamma$.
The resulting synthesis-based reweighted inpainting is summarized in Alg.\,\ref{alg:reweighting:synthesis}.
%, which is taken from \cite{Weinstein2011:DeclippingSparseland}.
Note that step 4 of the algorithm represents the weighted synthesis audio inpainting, and
therefore this step is carried out by Alg.\,\ref{alg:douglasrachford}.

%Alternatively,
Naturally,
the idea of reweighting can be included in the analysis-based recovery,
which was proposed already in \cite{CandesWakinBoyd2008:enhanced.reweighted.l1},
but not presented in the field of audio restoration.
%application to audio declipping in \cite{Weinstein2011:DeclippingSparseland}.
%The main difference to the synthesis model is that the solution
%of the analysis inpainting formulation
%is the signal in time domain,
%As a~consequence,
%thus
In contrast to the synthesis case, 
the analysis-based algorithm requires an additional application of the analysis operator,
in order to travel from the signal space
to the coefficient domain and thus to be able to assign the weights.
The algorithm is summarized in Alg.\,\ref{alg:reweighting:analysis}.
This time, step 4 is solved by Alg.\,\ref{alg:chambollepock}.

\vspace{-0.5em}
\begin{algorithm}%[h]
	\DontPrintSemicolon
	\SetKwInput{KwRequire}{require}
	\SetKw{KwReturn}{return}
	\SetKw{KwOR}{or}		
	\KwRequire{tight synthesis operator $\syn\colon\CC^N\to\CC^L$, set of feasible solutions $\Gamma\subset\CC^L$, parameters $K,\,\varepsilon,\,\delta>0$}
	set iteration counter $k = 1$\\
	set initial weights $w^{(1)}_i = 1,\,i = 1,\dots,N$\\
	\Repeat{$k > K$ \KwOR $\norm{\z^{(k)}-\z^{(k-1)}}_2<\delta$}
	{
		$\z^{(k)} = \argmin_\z \norm{\w^{(k)}\odot\z}_1\text{ s. t. }\syn\z\in\Gamma$\\
		$w^{(k)}_i = 1/(|z^{(k)}_i|+\varepsilon),\,i = 1,\dots,N$\\
		$k\leftarrow k+1$
	}
	\KwReturn{$\x = \syn\z^{(k-1)}$}
	\caption{Synthesis reweighted $\ell_1$ for inpainting}
	\label{alg:reweighting:synthesis}
\end{algorithm}
\vspace{-0.5em}

\begin{algorithm}%[h]
	\DontPrintSemicolon
	\SetKwInput{KwRequire}{require}
	\SetKw{KwReturn}{return}
	\SetKw{KwOR}{or}		
	\KwRequire{tight synthesis operator $\syn\colon\CC^N\to\CC^L$, set of feasible solutions $\Gamma\subset\CC^L$, parameters $K,\,\varepsilon,\,\delta>0$}
	set iteration counter $k = 1$\\
	set initial weights $w^{(1)}_i = 1,\,i = 1,\dots,N$\\
	\Repeat{$k > K$ \KwOR $\norm{\z^{(k)}-\z^{(k-1)}}_2<\delta$}
	{
		$\x^{(k)} = \argmin_\x \norm{\w^{(k)}\odot\ana\x}_1\text{ s. t. }\x\in\Gamma$\\
		$\z^{(k)} = \ana\x^{(k)}$\\
		$w^{(k)}_i = 1/(|z^{(k)}_i|+\varepsilon),\,i = 1,\dots,N$\\
		$k\leftarrow k+1$
	}
	\KwReturn{$\x^{(k-1)}$}
	\caption{Analysis reweighted $\ell_1$ for inpainting}
	\label{alg:reweighting:analysis}
\end{algorithm}
\vspace{-0.5em}

\section{Offset: possibly a strong influencer}
\label{sec:offset}
\begin{figure}[htb]
	\centering
	\subfloat[no offset]{\includegraphics[width=0.8\linewidth]{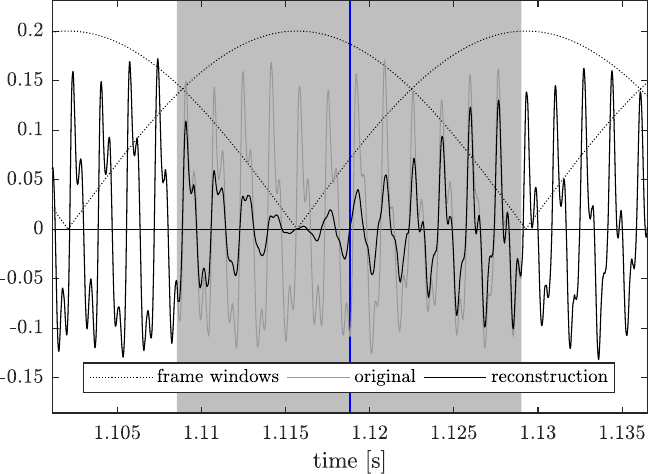}
		\label{fig:offset_intro_none}}
	\\
	\subfloat[full offset]{\includegraphics[width=0.8\linewidth]{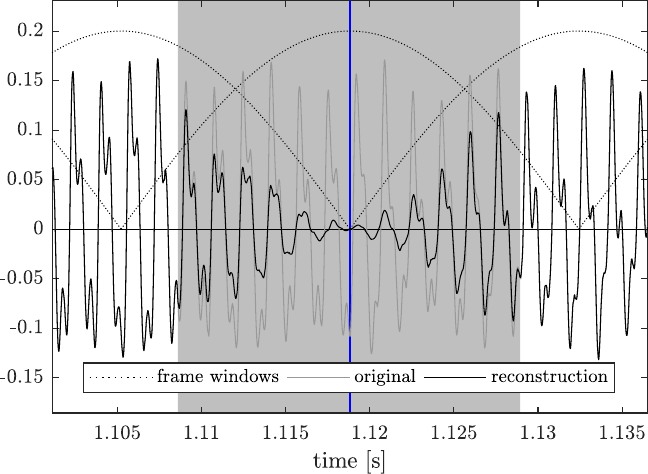}%
		\label{fig:offset_intro_full}}
	\caption{Visualization of energy drop in the gap and the effect of the offset. The center of the gap is denoted by the blue line.}
	\label{fig:offset_intro}
\end{figure}

The reader can see from Fig.\,\ref{fig:energy_decrease} that the minimum of the amplitude in the reconstructed signal does not appear exactly in the center of the gap.
The positioning of the Gabor system with respect to the location of the gap plays a role here.
In Fig.\,\ref{fig:offset_intro},
%two Gabor systems and their respective $\ell_1$ reconstructions 
$\ell_1$ reconstructions using two different Gabor systems
are presented.
The second system has been shifted
%with respect to the gap
such that the \qm{central} Gabor window fits the center of the gap.
Consequently, the energy of the reconstructed signal decreases symmetrically within the gap.

\begin{figure}[tbh]
	\centering
	\subfloat[full offset]{\includegraphics[width=0.48\linewidth]{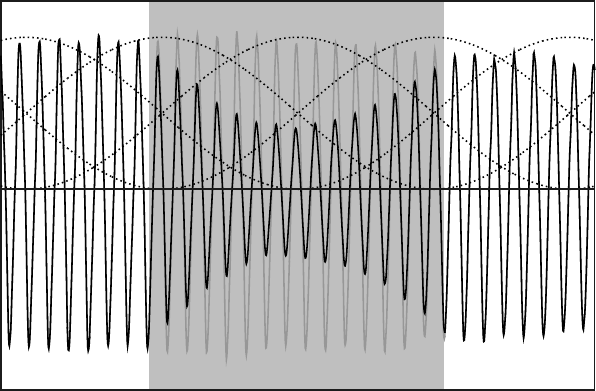}
		\label{fig:offset:full}}
	\hfill
	\subfloat[half offset]{\includegraphics[width=0.48\linewidth]{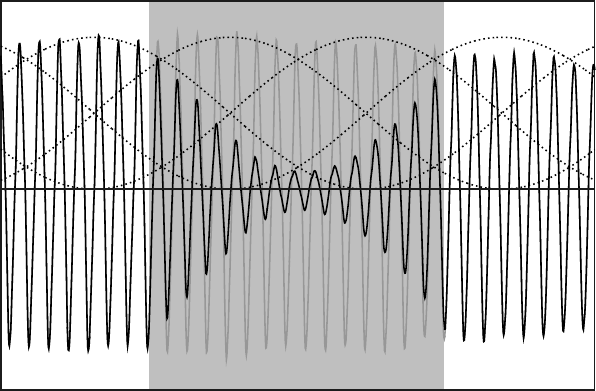}%
		\label{fig:offsete:half}}
	\caption{The two offset settings and their influence on the signal energy inside the filled gap.
		Original and restored signals are shown together with the individual shifts of Gabor windows.} % for both offset settings.}
	\label{fig:offset}
\end{figure}

We call the amount of the shift of the system \emph{the offset}.
We need to centralize the energy loss since methods compensating it in the time domain (Sections~\ref{sec:gradual} and \ref{sec:tdc}) require such a~symmetric setting.
%, we need this point of minimal energy to be in the middle of the filled segment.
%
%Up to this point, we did not make any specific assumptions about the location of minimal energy within the filled gap.
%In order to propose methods to compensate for the energy loss in time domain in Sec.\,\ref{sec:gradual} and \ref{sec:tdc}, we need this point of minimal energy to be in the middle of the filled segment.
%The proposed pre-processing method is based on the observation that the energy decrease is symmetric with respect to the central index of the gap, if the (shifted) frame windows are also symmetric in such way (see Fig.\,\ref{fig:offset_intro}).
%
In this section, we will propose two approaches---\emph{half} and \emph{full} offset---that will ensure the energy drop will be symmetric after running the inpainting algorithm described in Sec.\,\ref{sec:synthesis_analysis}.
There are two ways of choosing the offset:
either the center of the gap corresponds to the center of a~Gabor window (we refer to this as the \emph{full offset})
or the gap center is just in the middle between two neighboring windows (\emph{half offset}),
see the illustrative sequence of Gabor windows in Fig.\,\ref{fig:offset}.
Note that when we refer to the center of the window with the index $k+1$, we mean the signal index $1+k\cdot a$
\cite{RajmicBartlovaPrusaHolighaus2015:Acceleration.support.restriction},
assuming that the indexes of the signal samples start from~1.
The selection of the $\offset$ value is formalized in %the following algorithm.
Alg.\,\ref{alg:offset}.

%\begin{algorithmic}[1]
%	\REQUIRE
%		indexes $s$ and $f$ of the first and the last missing sample within the original signal;
%		Gabor window shift $a$
%	\STATE find the central index of the gap, $c = \ceil{(s+f)/2}$
%	\STATE
%		compute $k = \floor{(c-1)/a}$
%	\IF{full offset}
%		\STATE $d = 1+k\cdot a$
%	\ENDIF
%	\IF{half offset}
%		\STATE $d = 1+k\cdot a~+ \ceil{a/2}$
%	\ENDIF
%	\RETURN $\offset = c - d$	
%\end{algorithmic}
\vspace{-0.5em}
\begin{algorithm}[h]
	\DontPrintSemicolon
	\SetKwInput{KwRequire}{require}
	\SetKw{KwReturn}{return}	
	\KwRequire{indexes $s$ and $f$ of the first and the last missing sample within the original signal,
		Gabor window shift $a$}
	compute the central index of the gap $c = \ceil{(s+f)/2}$\\
	$k = \floor{(c-1)/a}$\\
	\If{full offset}{$d = 1+k\cdot a$}
	\If{half offset}{$d = 1+k\cdot a~+ \ceil{a/2}$}	
	\KwReturn{$\offset = c - d$}
	\caption{Computing the $\offset$ value}
	\label{alg:offset}
\end{algorithm}
\vspace{-0.5em}
%Line 2 finds $k$, the index of \dod{window such that the index $c$ is between the central indices of this and the following window}, i.e.\ $k \in \NN:1+k\cdot a\leq c \leq 1+(k+1)\cdot a$.
On line 2, the index of the nearest window preceding the index $c$ is found,
i.e.\ $k \in \NN:1+k\cdot a\leq c \leq 1+(k+1)\cdot a$.
%\todo{To je potřeba vyjasnit/opravit. Dokonce se v algoritmu ani nevyskytuje to $c$, až na výstupu.}
In the audio inpainting framework, the $\offset$ value is used as an input of the algorithm for support restriction %, presented in
\cite{RajmicBartlovaPrusaHolighaus2015:Acceleration.support.restriction}.

Fig.\,\ref{fig:offset} shows that the inclusion of one of the two offset choices in the $\ell_1$ inpainting algorithm ensures a~symmetric energy drop.
%with respect to the center of the gap.
We observe in this simulated example that the energy drop differs substantially for the two choices.

Does the full offset systematically provide better reconstruction than the half offset?
Better performance is \emph{on average} reached with the half offset.
Sec.\,\ref{sec:exp} will show an in-depth analysis and the dependence of the answer on the model and on whether the $\ell_1$ norm is weighted.

\section{Gradual inpainting}
\label{sec:gradual}
%\dod{může být jako sekce čtyři?}

Sec.\,\ref{sec:synthesis_analysis} proposed a~method that used weighting the signal coefficients to reduce the energy drop within the filled gap.
Sections \ref{sec:gradual} and \ref{sec:tdc} propose different methods for compensating for the same artifact,
but they are, by contrast, based on processing directly in the time domain.

%\subsection{Motivation}
%
%The experiments with the basic inpainting algorithms described in Sections \ref{sec:synthesis_analysis} and \ref{sec:offset} show that usually, as a~consequence of the $\ell_1$ relaxation, the reconstruction has lower energy in the filled part, as was described in Section \ref{sec:atom_weighting}. The energy is lowest in the center of the filled gap while the reconstruction on the borders of the filled gap is closer to the original signal (see Fig.\,\ref{fig:energy_decrease}).
%
%In Section \ref{sec:synthesis_analysis}, we proposed method to reduce this energy loss by weighting the frame atoms. The goal of sections \ref{sec:gradual}, and \ref{sec:tdc} is to propose different methods to compensate for the energy loss, which are, in contrast, based on the processing in time domain.

%\subsection{Main idea}

As can be seen in the figures presented,
%Fig.\,\ref{fig:energy_decrease},
the closer to the borders of the gap we are, the better the reconstruction is.
The \emph{gradual $\ell_1$ inpainting} starts from this observation:
Each time the signal in the gap is restored by $\ell_1$ minimization,
small chunks of the just computed samples from the beginning and from the end of the filled gap are fixed and treated as reliable from this moment on.
%for the next \qm{grade}.
At the next grade, inpainting is performed on a~gap that is accordingly shorter.
The process is repeated until the whole gap is filled,
%The visualization of the idea is in Fig.\,\ref{fig:gradual_idea}.
%
%\begin{figure}[htb]
%	\centering
%	\includegraphics[width=0.9\linewidth]{figures/gradual.pdf}
%	\caption{Illustration of the main idea of gradual inpainting.}
%	\label{fig:gradual_idea}
%\end{figure}
%
%\subsection{Algorithm}
%
which is formalized in %the following algorithm.
Alg.\,\ref{alg:gradual}.
Step 4 can utilize any of the above-presented signal models/algorithms.

%\begin{algorithmic}[1]
%	\REQUIRE
%		degraded signal $\y$;
%		indexes $s$ and $f$ of the first and the last missing sample, respectively
%	\STATE set grade counter $\grade = 0$
%	\STATE set $\y^{(0)} = \y$
%	\WHILE{$s \leq f$}
%		\STATE find $\y^{(\grade+1)}$ as the solution of inpainting problem with degraded signal $\y^{(\grade)}$, with samples
%			missing from $s$ to $f$
%		\STATE set step parameter $r\in \NN$ 
%		\STATE $s \leftarrow s + r$\quad \COMMENT{shrink the gap from left}
%		\STATE $f \leftarrow f - r$\quad \COMMENT{shrink the gap from right}
%		\STATE $\grade \leftarrow \grade +1$
%	\ENDWHILE
%	\RETURN $\y^{(\grade)}$
%\end{algorithmic}
\vspace{-0.5em}
\begin{algorithm}%[h]
	\DontPrintSemicolon
	\SetKwInput{KwRequire}{require}
	\SetKw{KwReturn}{return}	
	\KwRequire{degraded signal $\y$,
		indexes $s$ and $f$ of the first and the last missing sample, respectively}
	set grade counter $\grade = 0$\\
	set $\y^{(0)} = \y$\\
	\While{$s \leq f$}{
		find $\y^{(\grade+1)}$ as the solution to inpainting problem with degraded signal $\y^{(\grade)}$, with samples
		missing from $s$ to $f$\\
		set step parameter $r\in \NN$\\
		$s \leftarrow s + r$%\tcp*{shrink the gap from left}
		\tcp*{shrink from the left}
		$f \leftarrow f - r$%\tcp*{shrink the gap from right}
		\tcp*{shrink from the right}
		$\grade \leftarrow \grade +1$
	}
	\KwReturn{$\y^{(\grade)}$}
	\caption{Gradual inpainting}
	\label{alg:gradual}
\end{algorithm}
\vspace{-0.5em}
%\subsection{Weighting the atoms}

Using the notation 
%Regarding the described gradual algorithm and the notation
in problems \eqref{eq:inpainting_syn_unconstrained} and \eqref{eq:inpainting_ana_unconstrained},
this algorithm produces a~hierarchy of feasible solutions $\Gamma^{(\grade)} \subset \Gamma^{(\grade-1)}$
and can use potentially different weighting vectors $\w^{(\grade)}$
%for the inpainting problem
at each grade.
Clearly, the solution computed at grade $\grade$ satisfies
\begin{equation}
	\y^{(\grade)} \in \Gamma^{(\grade+1)}.
	\label{eq:iteration}
\end{equation}
Note that if no weighting is used at all, i.e.\ $\w^{(\grade)}\odot\x = \x$ at all grades $\grade$,
% \in \NN$,
the same objective function is minimized at each grade,
and \eqref{eq:iteration} induces $\y^{(\grade+1)} = \y^{(\grade)}$.
Therefore, the gradual approach with no weights does not find a~solution any different from the all-at-once approach.
The proposed weighting of the atoms thus forms one of the possible approaches, allowing the gradual algorithm to lead to a different (and possibly better) solution to the inpainting problem.

%\section{Direct time domain compensation for energy loss}
\section{Time domain compensation for energy loss}
\label{sec:tdc}

In this section, we propose a heuristic method for compensating for energy loss
 %within the gap
after running the $\ell_1$ minimization.
%-based inpainting algorithm.
The idea is to take the solution %to the $\ell_1$ minimization
%Instead of improving the algorithm, we
and modify it by entrywise multiplication of the recovered gap in the time domain by a~compensation curve, in order to increase its amplitude.
%i.e.\ its energy.
%This way, the energy drop would be \todo{anihilated}.
% The goal now is to propose an algorithm finding suitable compensation curve.

\subsection{Notation and requirements}
\label{sec:notation}
To formulate the demands on the compensation curve, assume for the moment that the signal is a~function on the interval $\langle 0,T\rangle$
% (instead of a vector)
with the gap spreading across $\langle s,f\rangle\subset\langle 0,T\rangle$.

Denote $c = (s+f)/2$ the center of the gap.
A %compensation
curve $q(t)$ is suitable for the energy loss compensation, if it satisfies the following natural conditions:
%%
%\begin{flalign*}
	%\quad\bullet
	%\quad	&q(t) \text{ is smooth on } \langle 0,T \rangle,&&\\
	%\bullet
	%&\frac{\mathrm{d}q}{\mathrm{d}t} = 0\hspace{0.35em}\quad\text{for}\quad t\in\{s,f\},&&\\[1mm]
	%\bullet
	%&q(t) = 1\quad\text{for}\quad t\notin\langle s,f\rangle,&&\\[2mm]
	%\bullet
	%&q(t)\text{ is non-decreasing on }(s,c),\text{ non-increasing on }(c,f),&&\\[2mm]
	%\bullet
	%&q(s+t) = q(f-t)\quad\text{for}\quad t\in\langle 0,f-s\rangle.
%\end{flalign*}
%%
%
\begin{itemize}
	\item
		$q(t)$ is smooth on $\langle 0,T \rangle,$
	\item
		$\frac{\mathrm{d}q}{\mathrm{d}t} = 0\hspace{0.35em}\quad\text{for}\quad t\in\{s,f\},$
	\item
		$q(t) = 1\quad\text{for}\quad t\notin\langle s,f\rangle$,
	\item
		$q(t)$ is non-decreasing on $(s,c)$, non-increasing on $(c,f)$,
	\item
		$q(s+t) = q(f-t)\quad\text{for}\quad t\in\langle 0,f-s\rangle$.
\end{itemize}
The first three conditions ensure that the adjustment of amplitude inside the filled gap is smooth.
The last two conditions reflect the observation that the greatest energy drop is in the center of the filled gap and it is symmetric.
In the discrete setting, the compensation vector $\q$ of length $h$ is obtained by sampling $q(t)$ in the interval $\langle s,f\rangle$.
% Since our problem setting is discrete, we need the compensation curve to be a vector of length $h$.
% Such vector $\q$ is obtained by discretizing the function  $q$ in the interval $\langle s,f\rangle$.
% The solution with compensated energy loss is then obtained by multiplying the filled segment entrywise by $\q$.

\subsection{Computing the compensation curve}
% The goal of this section is to find $q(t)$ respecting the conditions from Sec.\,\ref{sec:notation}, such that it also depends on the specific signal and context of the gap.
% The main idea is to use information from the (reliable) neighborhood of the gap.
%The goal of this section is to compute
Finding a~good heuristic curve $q(t)$ must be based on reliable information in the neighborhood of the gap.
Our approach assumes that in the neighborhood  of the gap, signal characteristics do not change too dynamically.
Additional gaps are artificially created and inpainted, which provides perfect local information about the energy decrease, since the reference original signal in these artificial gaps is available.
This information is then used to compensate for the energy loss in the gap that was originally treated.

\begin{figure*}[h]
	\centering
	\begin{subfloat}[the segmentwise energy progression in the original signal (black) and in the filled gap (blue)]{
		\includegraphics[width=0.40\textwidth]{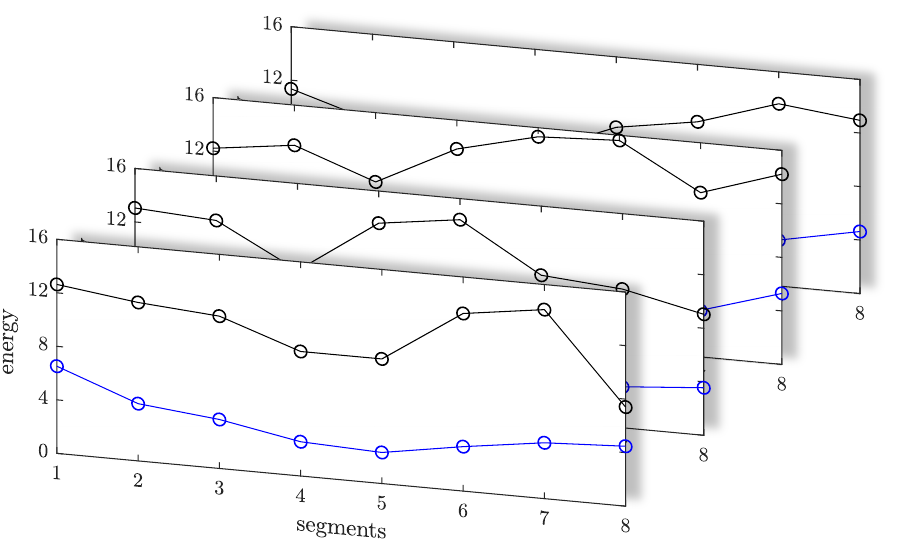}
		\label{fig:tdc_algo:energie_vse}}
	\end{subfloat}
	\hspace{0.05\textwidth}
	\begin{subfloat}[the ratio of the energy of the original signal and the filled gap through the segments]{
		\includegraphics[width=0.40\textwidth]{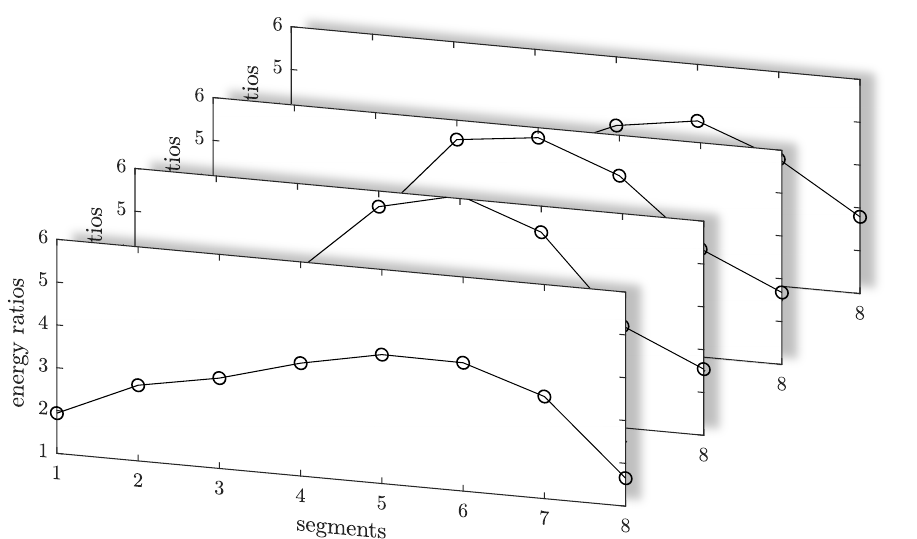}
		\label{fig:tdc_algo:pomery_vse}}
	\end{subfloat}
	\begin{subfloat}[the individual ratios of the energy together with the vector $\vect{m}$ computed as in Eq.\,\eqref{eq:multipliers}]{
		\includegraphics[width=0.40\textwidth]{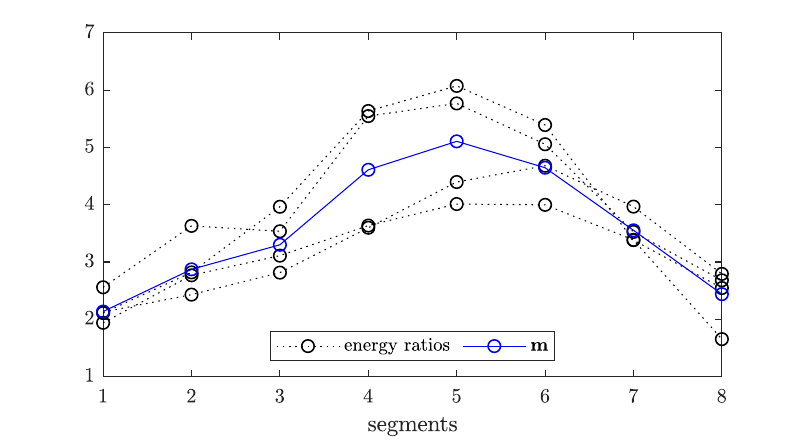}
		\label{fig:tdc_algo:pomery_a_m}}
	\end{subfloat}
	\hspace{0.05\textwidth}
	\begin{subfloat}[the symmetrization of the vector $\vect{n}=\sqrt{\vect{m}}$ and the resulting vector $\vect{q}$ found by spline interpolation]{
		\includegraphics[width=0.40\textwidth]{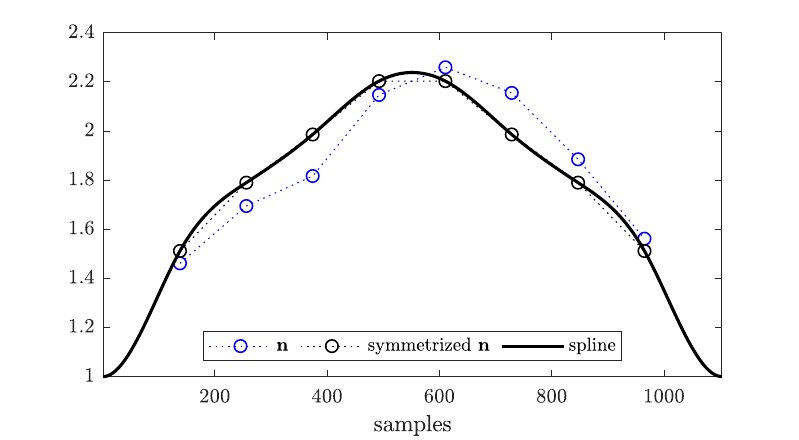}
		\label{fig:tdc_algo:spline}}
	\end{subfloat}
	\caption{Visualization of the direct time domain compensation for energy loss on a typical signal excerpt.
	For all the plots (i.e.\ for 4 additional new gaps), the energy is computed in $m = 8$ overlapping segments.}
	\label{fig:tdc_algo}
\end{figure*}

Below, we formalize the described idea.
See also Fig.\,\ref{fig:tdc_algo}, which illustrates the main steps of the algorithm.

\begin{figure*}
	\centering
	\includegraphics[width=0.8\linewidth]{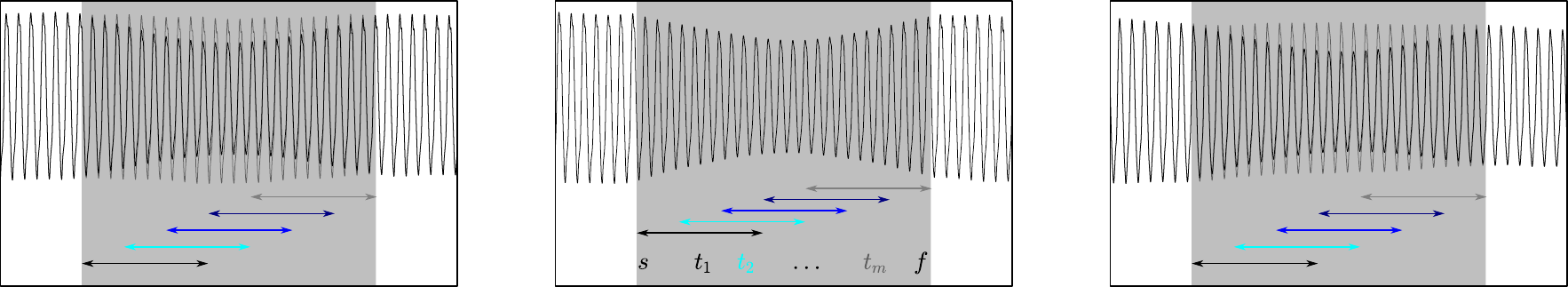}
	\caption{Illustration of how the segments are distributed inside the gap to compute the energy progression.
	For the original gap (middle plot), two artificial gaps are created (left and right), where the reference signal (gray) is available.
	The distribution of $m = 5$ segments, represented by the arrows, is the same in all three cases.}
	\label{fig:segments}
\end{figure*}

\begin{enumerate}
	\item In the (reliable) neighborhood of the gap, create new gaps.
	\item Inpaint the original as well as the additional gaps, using the same setting.
	% \item In all the new gaps \todo{and the corresponding parts of the reliable signal},
	% compute how the energy progresses through the gap.
	\item For all the new gaps, compute how the energy progresses through the gap and through the corresponding portion of the reliable signal. 
	The energy progression is estimated via $m$ overlapping signal segments covering the whole inpainted gap.
	For the following steps, the segments should be distributed for all the gaps \emph{in the same way}, such that the information is transferable into the time instants
	% These segments are centered around time instants
	$s < t_1 < \dots < t_m < f$
	in the original gap (see Fig.\,\ref{fig:segments}).
	% The energy progression is thus stored in vectors of length $m$.
	\item Find the multipliers $\vect{m}\in\RR^m$, such that the difference in energy between the original signal and the filled parts is minimized (see Eq.\,\eqref{eq:multipliers}).
	\item Compute $\vect{n} = \sqrt{\vect{m}}$ using the entrywise square root\footnote{The vector $\vect{m}$ represents the ratios of energy, while we need the ratio of signal amplitude. This corresponds to the square root of the ratios of energy.}.
	\item Enforce symmetry\footnote{We use the offset described in Sec.\,\ref{sec:offset}, which makes the assumption of symmetric energy drop realistic.} of $\vect{n}$ by updating
	\begin{subequations}
	\begin{align}
	n_i &\leftarrow \frac{n_i + n_{m+1-i}}{2}, &1&\leq i\leq \floor{\frac{m}{2}}\\
	n_i &\leftarrow n_{m+1-i}, &\floor{\frac{m}{2}}+1&\leq i\leq m .
	\end{align}
	\end{subequations}
	\item The function $q(t)$ is obtained by cubic spline interpolation%
	\footnote{We use the MATLAB function \texttt{spline}, see \url{https://www.mathworks.com/help/matlab/ref/spline.html}.},
	which for $t_1,t_2,\dots,t_m$ attains the values of the vector $\vect{n}$, and at points $s$ and~$f$ its value is 1 and derivative 0, respectively.
	\item The vector $\q$ is obtained from $q(t)$ by equidistant sampling in the interval $\langle s,f\rangle$.
\end{enumerate}

To clarify step 4, define two matrices.
Let the columns of the matrix $\matr{X}$ be formed by the vectors of energy progression from the additional gaps.
Let the matrix $\matr{Y}$ contain the respective values of energy from the original signal.
Note that this way, both matrices have $m$ rows and the number of columns is the number of additional gaps
(created in step~1).
The multipliers $\vect{m}$ are then the optimizers of 
\begin{equation}
\vect{m} = \argmin_{\vect{m}'\in\RR^m}\norm{\matr{Y}-\mathrm{diag}(\vect{m}')\cdot\matr{X}}_2^2.
\label{eq:multipliers}
\end{equation}
%
% From Eq.\,\eqref{eq:multipliers}, it is possible to interpret the vector $\vect{m}$.
To understand Eq.\,\eqref{eq:multipliers},
imagine a~column in $\matr{X}$ (the energy of the inpainted signal).
Multiplying it elementwise by $\vect{m}$,
we want to reach as close as possible to the corresponding column of $\matr{Y}$ (original signal).
The need for the minimization comes from the fact that the vector $\vect{m}$ is common to all the columns.
% The energy of the filled gap, multiplied by segments by $\vect{m}$, will in the context of the additional gaps have the energy as close as possible to the original signal.

Eq.\,\eqref{eq:multipliers} is a~least squares problem and its solution can be written explicitly using the entries of matrices $\matr{X}$ and $\matr{Y}$ as
\begin{equation}
m_i = \frac{\sum_j y_{ij}x_{ij}}{\sum_j x_{ij}^2},\quad i = 1,\dots,m,
\label{eq:multipliers_solution}
\end{equation}
which corresponds to a very fast procedure.

\section{Experiments}
\label{sec:exp}

This section presents a numerical evaluation of the above-described approaches to energy loss compensation.
The last experiment shows an overall comparison. %combines the methods together.

\subsection{Performance measures}

As the standard performance measure, we use the signal-to-noise ratio (\SNR), defined as
\begin{equation}
	\mathrm{SNR}\left(\y_{\text{orig}},\y_{\text{inp}}\right) = 10 \cdot\log_{10}\dfrac{\norm{ \y_{\text{orig}}}_2^2}{\norm{\y_{\text{orig}}-\y_{\text{inp}}}_2^2}\quad\text{[dB]},
	\label{eq:SNR}
\end{equation}
where $\y_{\text{inp}}$ stands for the recovered signal
%filled segment of missing samples
and $\y_{\text{orig}}$ denotes the original signal \cite{Adler2012:Audio.inpainting}.
%\footnote{We can utilize this formula since our signals under examination are not a~priori degraded.}.
%
%therefore for any gap we know the optimal result that we want to achieve.
%Note that it makes no difference if we compute the SNR on whole signals or we restrict to the gap,
%since in both synthesis and analysis models we force the consistency of the solutions in their reliable part
%(sets $\adjoint{\Gamma}$ and $\Gamma$, respectively).
Recall that the very last step of our reconstruction algorithms is the projection onto the set $\Gamma$.
In our implementation, we therefore evaluate the difference in the gap only, since elsewhere
$\y_{\text{orig}}-\y_{\text{inp}}$ contains solely zeros.
%since in both the synthesis and analysis model we add projection onto the set $\Gamma$ as final step of the inpainting algorithm, therefore the condition of equality of the original and restored signal in reliable parts is satisfied.
Obviously, higher \SNR{} values reflect better reconstruction.

The \SNR{} formula does not compensate for the length of the filled gap.
%: imagine two recovered gaps with the same character and degree of degradation, compared to the original signal---for example, the same frequency component is added to the original signal;
%the $\SNR$ of the longer gap will automatically be lower.
This is the reason why only gaps of identical lengths will be taken into account in our comparisons.
%,except for rare cases where we will show aggregate results.
%
%---for example reconstruction of very noisy signal should have the same energy as the original signal but the subjective impression does not depend very much of the spectrum of the reconstruction.
%\todo{moc nevim, cos tim chtel rict}
%
Note that we compute the average \SNR{} by first computing the particular values of \SNR{} in dB, and then taking the average.

%Also note that this objective measure often does not correspond to the subjective perception of the reconstructed signal.
For the final, overall comparison, we also included the \mbox{PEMO-Q} evaluator \cite{Huber:2006a}.
This tool provides an evaluation that takes the model of human auditory system into account, thus being closer to the subjective evaluation than the SNR.
%The results are thus closer to evaluation in listening tests while still being objective.
The measured quantity
% are \emph{perceptual similarity measure} (specifically 5th percentile of the time-dependent values, \PSMt)
called \emph{objective difference grade} (\ODG) can be interpreted as
the degree of perceptual similarity between $\y_{\text{orig}}$ and $\y_{\text{inp}}$.
%The values of \PSMt{} are between 0 and 1 (higher value represents higher similarity).
The \ODG{} attains values from $-4$ (very annoying) up to $0$ (imperceptible), reflecting the effect of audio artifacts in the restored signal.
%and the interpretation is the following: \todo{možné zkrátit a dát na řádek}
%\begin{flalign*}
%0\quad&\text{imperceptible,}&&\\
%-1\quad&\text{perceptible, but not annoying,}&&\\
%-2\quad&\text{slightly annoying,}&&\\
%-3\quad&\text{annoying,}&&\\
%-4\quad&\text{very annoying.}&&
%\end{flalign*}

\editt{%
	The \ODG{} is an approximation of the subjective evaluation of restoration quality.
	Performing a~listening test, however, is intractable in the case of our experiment.
	The reasons are the extensive number of combinations and the non-existence of a~standardized procedure for evaluating the inpainted signals.
	For individual subjective evaluation, the reader is kindly asked to visit the accompanying web page
	(see the link in Sec.\,\ref{sec:soft}), where the restored signals can be played back interactively.%
}

\subsection{Evaluation setup}
We use a collection of ten music recordings sampled at 44.1\,kHz, with different levels of sparsity with respect to the Gabor representation.
Our signals were chosen from the EBU SQAM dataset \cite{EBUSQAM} \editt{and shortened to approximately 7\,s long excerpts.}
In each test instance, the input was a~signal containing 8 gaps at random positions.
The lengths of the gaps ranged from 5\,ms up to 50\,ms.
%

%\subsection{Used transform}
As the default choice in the 
%following 
tests, we used a~tight Gabor frame with the Hann window of length $w = 2800$ samples (approximately 64\,ms),
window shift $a = 700$ samples and with $M = 2800$ frequency channels.

%\begin{figure*}[h]
%	\centering
%	\begin{subfloat}[]{
%			\includegraphics[width=0.40\textwidth]{}
%			\label{}
%	\end{subfloat}
%	\hspace{0.05\textwidth}
%	\begin{subfloat}[]{
%			\includegraphics[width=0.40\textwidth]{}
%			\label{}}
%	\end{subfloat}
%	\caption{}\label{}
%\end{figure*}

%\subsection{Algorithms}
The Douglas--Rachford (DR) and Chambolle--Pock (CP) algorithms are used for the signal recovery,
%, as introduced in
see Sections \ref{Sec:Solving.synthesis.problem} and \ref{Sec:Solving.analysis.problem}, respectively.
\edit{The DR algorithm uses $\tau=0.2$, the CP algorithm uses $\tau=0.2$ and $\sigma=5$.
Iterations are terminated if the proposed criterion with $\varepsilon=5\cdot10^{-4}$ is satisfied or alternatively after $500$ iterations.}

For the sake of the overall comparison in Sec.\,\ref{sec:exp_overall},
the frame-wise Janssen algorithm \cite{javevr86}, which is based on linear prediction, was also included,
as used in \cite{Adler2012:Audio.inpainting}.
\edit{Furthermore, the precursor of sparsity-based methods, the Orthogonal Mathching Pursuit (OMP),
%is taken into account \cite{Adler2012:Audio.inpainting}.}
is included \cite{Adler2012:Audio.inpainting}.}
As the last competitor, we chose the SPAIN algorithm both in its synthesis and in its analysis variant
\cite{MokryZaviskaRajmicVesely2019:SPAIN}.
SPAIN used the same window and overlap as the Gabor transform did.

\subsection{\edit{Weighting the atoms and offset}}
\label{sec:exp_weighting}

\begin{figure}
	\centering
	\includegraphics[width=0.85\linewidth]{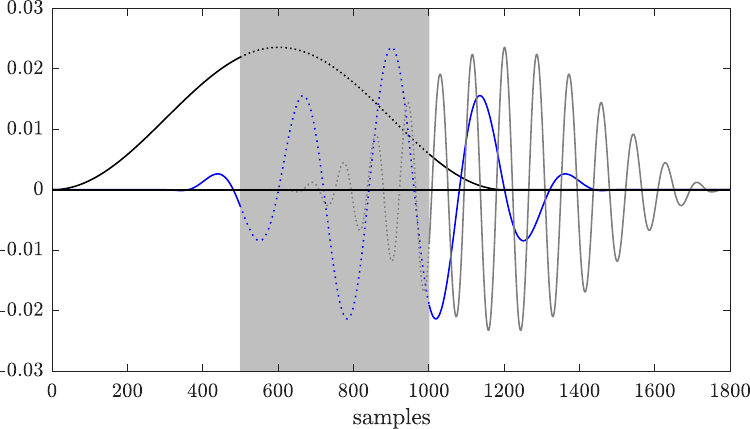}
	\caption{Three examples of Gabor atoms $\vect{d}_n$ with different modulations, overlapping with the gap.
		Only their real parts are depicted.
		%Degraded atoms, three settings of shift and modulation.
		Solid lines indicate their reliable part $\Mr\vect{d}_n$.}
	\label{fig:degraded_atoms}
\end{figure}

\edit{The motivation for weighting the atoms of the Gabor frame was presented in Section \ref{sec:atom_weighting}.
Now we comment in detail on the choices of the weights.}
%the frame atoms, as mentioned in Section \ref{sec:atom_weighting}.
Let $\vect{d}_n$ be
%the $i$-th atom of a~Gabor dictionary 
a~Gabor atom and let $\Mr\vect{d}_n$ be its part corresponding to the reliable part of the signal, see Fig.\,\ref{fig:degraded_atoms}.
We propose five different formulas for choosing the weights $\w$, including the vector of ones representing the non-weighted case.
The proposed formulas are arranged according to the variance of obtained weights, in ascending order.
\begin{tabbing}
	\hspace{2mm}\=\hspace*{10mm}\=\hspace*{35mm}\=\hspace*{25mm}\=\kill
	\>(a) \> $w_{n} = 1 $ \> no weighting \> (\texttt{none}) \\[3mm]
	\>(b) \> $w_{n} = \dfrac{|\supp(\Mr\vect{d}_n)|}{|\supp(\vect{d}_n)|}$ \> support-based \> (\texttt{supp}) \\[2mm]
	\>(c) \> $w_{n} = \dfrac{\norm{\Mr\vect{d}_n}_1}{\norm{\vect{d}_n}_1}$ \> $\ell_1$ norm-based \> (\texttt{abs}) \\[2mm]
	\>(d) \> $w_{n} = \dfrac{\norm{\Mr\vect{d}_n}_2}{\norm{\vect{d}_n}_2}$ \> $\ell_2$ norm-based \> (\texttt{norm}) \\[2mm]
	\>(e) \> $w_{n} = \dfrac{\norm{\Mr\vect{d}_n}_2^2}{\norm{\vect{d}_n}_2^2}$ \> energy-based \> (\texttt{energy})
\end{tabbing}
Note that \cite{Adler2012:Audio.inpainting} used the weighting based on the $\ell_2$ norm
in their synthesis model, such that the $\ell_2$ norms of atoms $\Mr\vect{d}_n$ were made identical.
This corresponds to our case (d).
The difference from \cite{Adler2012:Audio.inpainting} is that we use the $\ell_1$ approach instead of the greedy solver.
%as in \cite{Adler2012:Audio.inpainting}.

%\todo{je potřeba tento odstavec? pokud ano, přepsat srozumitelněji} Note that in the case of Gabor frames, the original atoms,
%i.e.\ $\vect{d}_n$ as translations and modulations of $\g$ have \dod{equal absolute values}.
%Therefore the denominators are not needed in the above formulas,
%nevertheless they support the understanding of our motivation for such weighting.

\begin{figure*}[t]
	\centering
	\includegraphics[width=\linewidth]{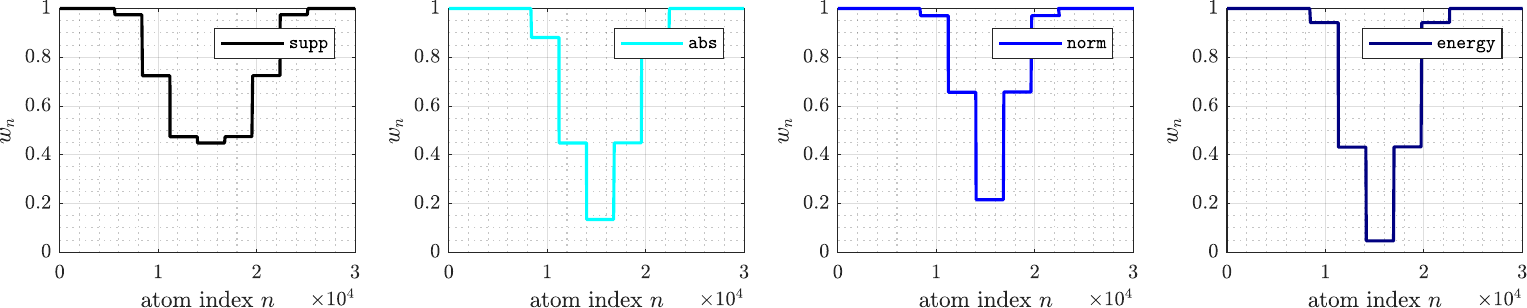}
	\caption{Illustration of the weighting variants (besides the obvious case of constant weights).
		The weights were computed using the Gabor frame with a 64\,ms long Hann window, overlap 75\,\% and the number of frequency channels equal to the length of the window in samples.
		The gap length was 35\,ms.
		%The piecewise character corresponds to the fact that each time-shift of the window has a~number of modulations.%
	}
	\label{fig:vahy}
	\vspace{-0.5em}
\end{figure*}

Fig.\,\ref{fig:vahy} plots the values of $\w$ for the proposed methods (b)--(e) for a~fixed gap length.
It illustrates that %these are symmetric (in line with the used offset method) and that
different options provide different weights.
\edit{%
	Furthermore, observe that all the values are strictly greater than zero, which follows from the window length being greater than the gap length.
	Should the opposite situation occur, some weights could be set to zero using any of the formulas (b)--(e).
	This would result in some coefficients not being penalized at all, while not being bound by the reliable signal parts either.
	In the iterative solver, the outcome would be that these coefficients would 
	keep their initial value.
	%remain unchanged \todo{in the course of iterations, meaning that their initial value would play a~role}.
}

Before proceeding to the performance evaluation based on choosing the weighting types, we analyze the effect of the offset (Sec.\,\ref{sec:offset}).
The motivation is that if a preferable offset option is found, then this type of offset will be used in all of the experiments focusing on other parameters.
Recall that Figures \ref{fig:offset_intro} and~\ref{fig:offset} illustrate
not only the connection between the value of offset and the position of minimal amplitude (inside the gap), but also the difference that the choice of offset variant makes.

\begin{figure*}[h]		
	\centering
	\begin{subfloat}[synthesis model]{
			\includegraphics[width=0.38\textwidth]{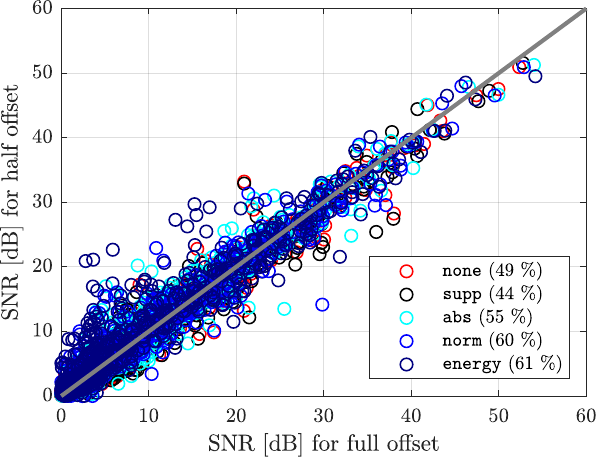}
			\label{fig:full_vs_half:DR}}
	\end{subfloat}
	\hspace{0.05\textwidth}
	\begin{subfloat}[analysis model]{
			\includegraphics[width=0.38\textwidth]{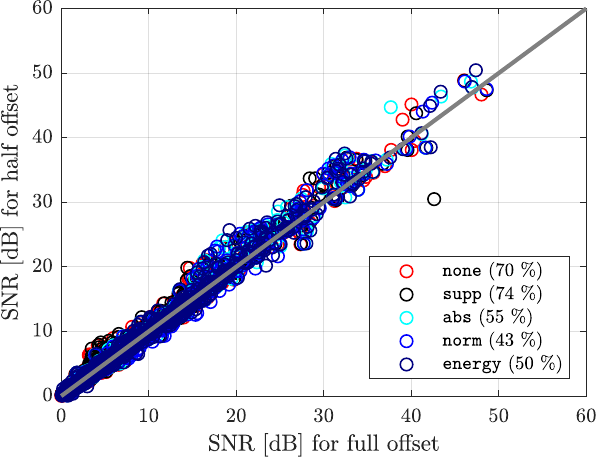}
			\label{fig:full_vs_half:CP}}
	\end{subfloat}
	\caption{%
		Comparison of the two offset approaches, combined with weighting the frame atoms. Every point represents a single test instance.
		For better orientation, the diagonal line is shown.
		\edit{The percentage in brackets represents the fraction of instances above the diagonal line, given the weighting type.}
	}
	\label{fig:full_vs_half}
	\vspace{-0.75em}
\end{figure*}

To analyze the difference between the choice of full or half offset,
the \SNR{} values for weighted $\ell_1$ inpainting are plotted in Fig.\,\ref{fig:full_vs_half}.
Two observations are clear from the figure.
First, there is no obvious way of choosing the offset value, since no clear dependence on the \SNR{} value or weighting type is observed.
However, the results slightly favor the half offset
(approx. 54\,\% of data points in Fig.\,\ref{fig:full_vs_half:DR} and 58\,\% in Fig.\,\ref{fig:full_vs_half:CP} lie above the diagonal line).
Second, the SNR values exhibit much more variation in the synthesis case, compared to the analysis case.
The two observations suggest that the choice of offset is not crucial in the analysis-based inpainting, whereas it could affect the results based on the synthesis model (due to the larger variance).

Since it is convenient to fix as many parameters as possible for the following comparisons,
and also the whole evaluation is based on an average performance of the algorithms, half offset will be the default choice in the subsequent experiments.
For the results with full offset, see the accompanying repository (link in Sec.\,\ref{sec:soft}).

\begin{figure*}[h]
	\centering
	\begin{subfloat}[synthesis model]{
			\includegraphics[width=0.38\textwidth]{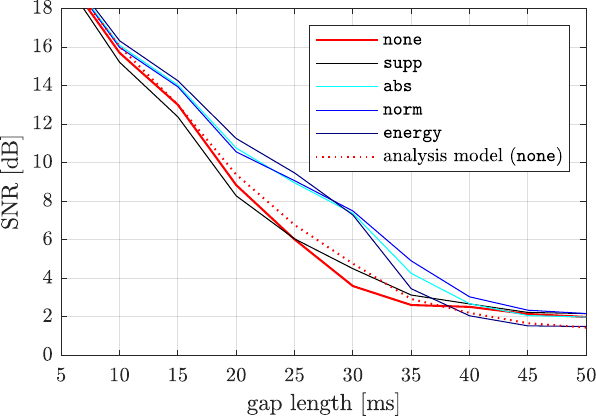}
			\label{fig:weighting:DR}}
	\end{subfloat}
	\hspace{0.05\textwidth}
	\begin{subfloat}[analysis model]{
			\includegraphics[width=0.38\textwidth]{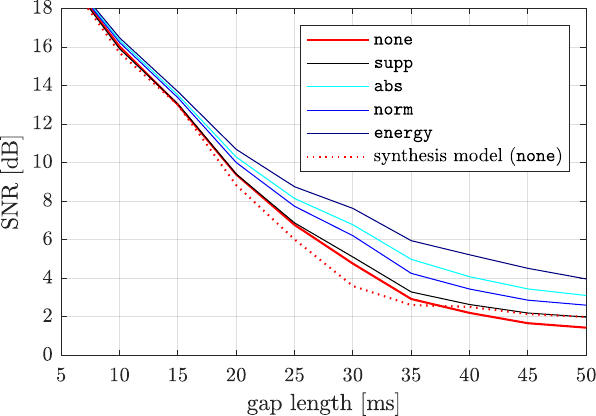}
			\label{fig:weighting:CP}}
	\end{subfloat}
	\caption{%
		Evaluation of the methods for weighting atoms.
		The simple (non-weighted) variant of the concurrent model is also depicted by dots as a reference.
	}
	\label{fig:weighting}
	\vspace{-0.5em}
\end{figure*}

Fig.\,\ref{fig:weighting} shows the influence of weighting (Sec.\,\ref{sec:atom_weighting}) on the results.
 %of inpainting.
 %an inpainting algorithm.
Regarding the basic, non-weighted models, one may observe that the analysis-based inpainting performs slightly better compared to the synthesis-based case, especially for the middle-length gaps.
It is, however, more interesting to notice that introducing the weights does not have the same effect in both models.
The performance of the analysis model improves as the variance of the involved weights grows
(see Fig.\,\ref{fig:vahy}).
In the synthesis model, on the other hand, weighting by the $\ell_1$ or $\ell_2$ norm leads to a~consistent improvement,
whereas the other choices may even decrease the value of \SNR.
Also note that in the synthesis model, the weighting is most beneficial for middle length gaps, while in the analysis model, the improvement increases with the gap length.

\subsection{Iterative reweighting}

Recall that in the iterative approach
%to weighting the frame atoms
(Sec.\,\ref{sec:reweighting}),
%is rather different compared to the proposed weighting based on atom overlap with the gap.
the iterative weights (denoted \texttt{iterative}) depend on both the signal
and the time-frequency transform, and that they are computed differently
in the synthesis- and analysis-based models.
In the analysis-based Alg.\,\ref{alg:reweighting:analysis},
the new weights are computed from coefficients in the range space of $\ana$,
which is generally not the case of the synthesis-based Alg.\,\ref{alg:reweighting:synthesis}.

\begin{figure*}[h]
	\centering
	\begin{subfloat}[synthesis model]{
			\includegraphics[width=0.39\textwidth]{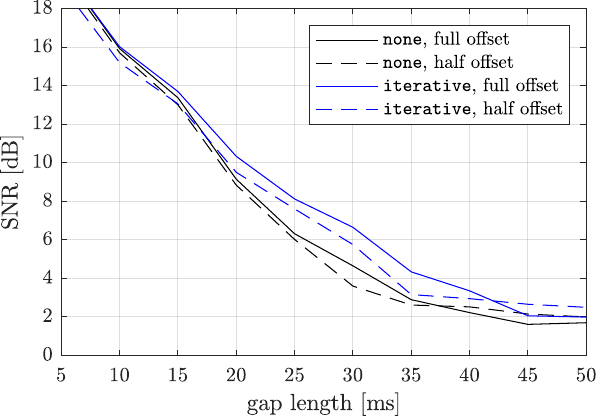}
			\label{fig:reweighting:DR}}
	\end{subfloat}
	\hspace{0.05\textwidth}
	\begin{subfloat}[analysis model]{
			\includegraphics[width=0.39\textwidth]{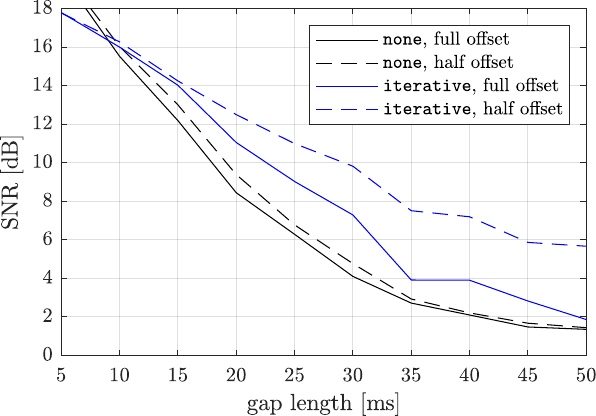}
			\label{fig:reweighting:CP}}
	\end{subfloat}
	\caption{%
		Evaluation of the methods for the iterative approach to weighting atoms.
		In this case, results for both the full and the half offset are shown.
		The parameters of the reweighting (see Alg.\,\ref{alg:reweighting:synthesis} and \ref{alg:reweighting:analysis}) are $K = 10$, $\varepsilon = 0.001$ and $\delta = 0.01$.
	}
	\label{fig:reweighting}
	\vspace{-0.75em}
\end{figure*}

For these reasons, Fig.\,\ref{fig:full_vs_half} and the earlier decision to use half offset are irrelevant in the case of iterative reweighting.
Fig.\,\ref{fig:reweighting} thus shows the average values of \SNR{} for both the half and the full offset,
in comparison with the simple non-weighted approach.
Fig.\,\ref{fig:reweighting:DR} shows that in the synthesis model, the iterative reweighting provides consistent
(however small)
improvement for longer gaps and it is not much dependent on the offset choice.
Contrarily, Fig.\,\ref{fig:reweighting:CP} suggests that the choice of offset is crucial in the reweighted analysis case.
Note that this result is in contrast to the observation for the non-iterative weighting,
where the choice of the offset did not significantly affect the results in the analysis case.
Note also that although the iterative reweighting is highly beneficial for longer gaps with half offset,
it decreases the average performance for the shortest gaps,
independently of the offset variant.

\subsection{Gradual inpainting}
\label{sec:exp_gradual}

\begin{figure*}[h]
	\centering
	\begin{subfloat}[synthesis model (\texttt{norm})]{%
			\includegraphics[width=0.325\textwidth]{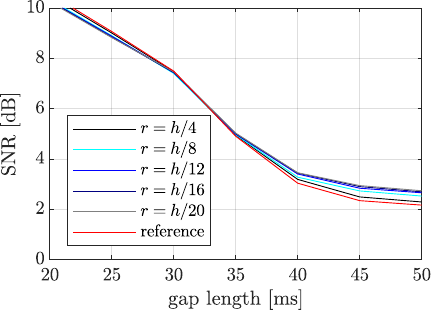}%
			\label{fig:gradual:DR}}%
	\end{subfloat}%
	\hfill%
	\begin{subfloat}[analysis model (\texttt{energy})]{%
			\includegraphics[width=0.325\textwidth]{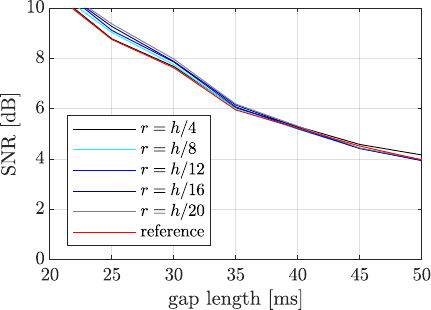}%
			\label{fig:gradual:CP}}%
	\end{subfloat}%
	\hfill%
	\begin{subfloat}[\edit{analysis model (\texttt{none})}]{%
			\includegraphics[width=0.325\textwidth]{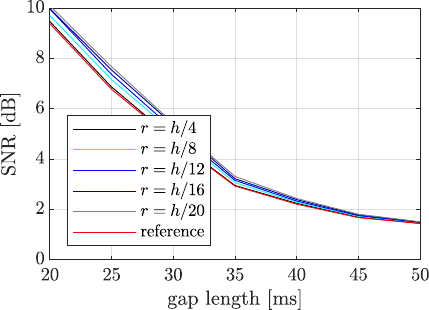}%
			\label{fig:gradual:CP_reference}}%
	\end{subfloat}%
	\caption{%
		Evaluation of the gradual approach.
		The results are plotted for different choices of the parameter $r$.
		The weighting option is chosen based on Figures \ref{fig:gradual:DR} and~\ref{fig:gradual:CP}.
		The non-gradual approach with the corresponding weighting is taken as the reference and plotted in red.
		}
		\label{fig:gradual}
		\vspace{-0.5em}
\end{figure*}

The average \SNR{} values for
%the experiment on
gradual inpainting introduced in Sec.\,\ref{sec:gradual} are presented in
Fig.\,\ref{fig:gradual}.
As mentioned above, the gradual approach needs to be % combined
fused with another modification of the $\ell_1$ inpainting to produce sensible results.
In this experiment, weighting the atoms was chosen with the weighting formula based on the results described in Sec.\,\ref{sec:exp_weighting}.
The only other parameter of the method is the number of samples $r$ taken as reliable from the left and right sides of the gap at each grade.

The results are presented for different choices of $r$ as a~fraction of the gap length $h$.
In the synthesis case, Fig.\,\ref{fig:gradual:DR} shows that the gradual algorithm is beneficial compared to the reference (the non-gradual method) for long gaps.
In the analysis case, on the other hand, Fig.\,\ref{fig:gradual:CP} clearly indicates that its performance via the gradual approach \edit{%
	does not improve.
	Fig.\,\ref{fig:gradual:CP_reference} illustrates that the variance in the results in Fig.\,\ref{fig:gradual:CP} is explained by the variance due to the approximate solution to the optimization problems involved.
	When any minor computational error appears, it is amplified at each grade of the gradual algorithm.
	Such a problem does not occur in the synthesis case, which was not expected.
}

%Nevertheless, in both cases, smaller values of $r$ result in a~larger difference in \SNR.
Finally, note that even with the positive effect of the gradual approach,
the synthesis model did not reach the quality  of the analysis model.

\subsection{Direct time domain compensation for energy loss}
\label{sec:exp_tdc}

The method of direct time domain compensation
 %for energy loss
(Sec.\,\ref{sec:tdc})
depends on a~larger number of parameters, compared to the previous techniques.
They are:
\begin{itemize}
	\item the number of additional artificial gaps (denoted \texttt{gaps}),
	\item the positions of these gaps in the signal,
	\item the number of segments from which the evolution of energy is computed (denoted $m$),
	\item the length of these segments.
\end{itemize}
In our experiment, 
%Testing all the combinations of the parameters would not lead to a simple conclusion, therefore
all the parameters except \texttt{gaps} have been fixed as follows:
\begin{itemize}
	\item the additional gaps are symmetrically distributed around the initial gap, starting $w$ samples away from the edge of the gap and then shifted by $w/2$ samples,
	\item the number of segments $m = 10$ and the length of each segment is $h/4$
			(i.e.\ the segments are overlapping).
\end{itemize}

\begin{figure*}[h]
	\centering
	\begin{subfloat}[synthesis model (\texttt{none})]{
		\includegraphics[width=0.37\textwidth]{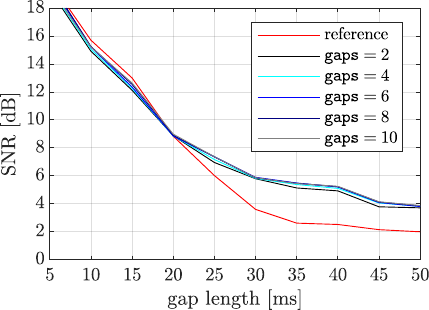}
		\label{fig:SNR:syn_none}}
	\end{subfloat}
	\hspace{0.05\textwidth}
	\begin{subfloat}[analysis model (\texttt{none})]{
		\includegraphics[width=0.37\textwidth]{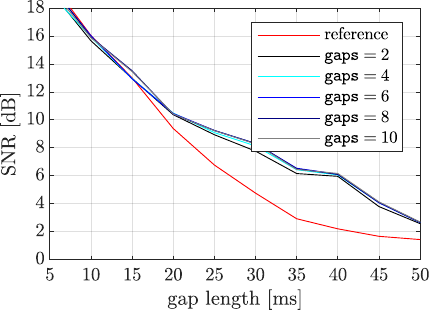}
		\label{fig:SNR:ana_none}}
	\end{subfloat}	
	\begin{subfloat}[synthesis model (\texttt{norm})]{
		\includegraphics[width=0.37\textwidth]{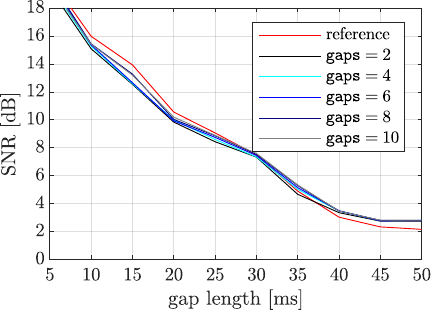}
		\label{fig:SNR:syn_norm}}
	\end{subfloat}
	\hspace{0.05\textwidth}
	\begin{subfloat}[analysis model (\texttt{energy})]{
		\includegraphics[width=0.37\textwidth]{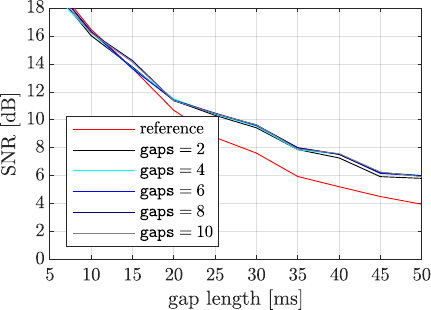}
		\label{fig:SNR:ana_energy}}
	\end{subfloat}

	\caption{%
		Evaluation of the direct time domain compensation for energy loss.
		The results are plotted for different choices of the number of additional gaps (denoted \texttt{gaps}).
		The reference uses the same weighting of atoms but no additional compensation method.
		The weighting for the plots \ref{fig:SNR:syn_norm} and \ref{fig:SNR:ana_energy} is based on the previous experiments (Fig.\,\ref{fig:weighting}).
	}
	\label{fig:SNR}
	\vspace{-0.5em}
\end{figure*}

Fig.\,\ref{fig:SNR} shows the results of the experiment.
Two approaches are tested in both the synthesis and the analysis model.
First, the time domain approach is meant to be a competitor of weighting the atoms in the coefficient domain
(plots \ref{fig:SNR:syn_none} and \ref{fig:SNR:ana_none}).
Second, the weighting in the coefficient and the time domains is combined to provide even more energy compensation (plots \ref{fig:SNR:syn_norm} and \ref{fig:SNR:ana_energy}).

One can conclude that except the synthesis model combined with weighting the atoms (Fig.\,\ref{fig:SNR:syn_norm}), the time domain compensation leads to an improvement by a~few dB.
The results are remarkable, especially when the analysis model is used.
Although Fig.\,\ref{fig:SNR:ana_none} shows that the time domain compensation does not surpass the weighting of the atoms
(compare with the reference in Fig.\,\ref{fig:SNR:ana_energy}),
the combination of both approaches proves to be the best choice.

Note also that all the plots suggest that the number of additional gaps does not crucially affect the results.

\subsection{Overall comparison}
\label{sec:exp_overall}

For the sake of an overall comparison,
%To specify the parameters of chosen methods for this comparison,
the time domain compensation for energy loss was applied with $\texttt{gaps} = 4$
while the SPAIN \edit{and OMP} algorithms used the frame-wise DFT dictionary with redundancy 4.
Finally, the frame-wise Janssen algorithm was applied \edit{according to \cite{Adler2012:Audio.inpainting}} with the order of the autoregressive model
%\begin{equation}
	$p = \min(3H+2,w/3)$,
%\end{equation}
%\todo{možné dát na řádek}
%\todo{zdůvodnění? např. že to stejně dělali v [9]?}
with $H$ denoting the number of missing samples within the current frame (window),
and the number of iterations was set to 50.

\begin{figure*}[h]
	\centering
	\begin{subfloat}[evaluation by \SNR]{%
			\includegraphics[width=0.47\textwidth]{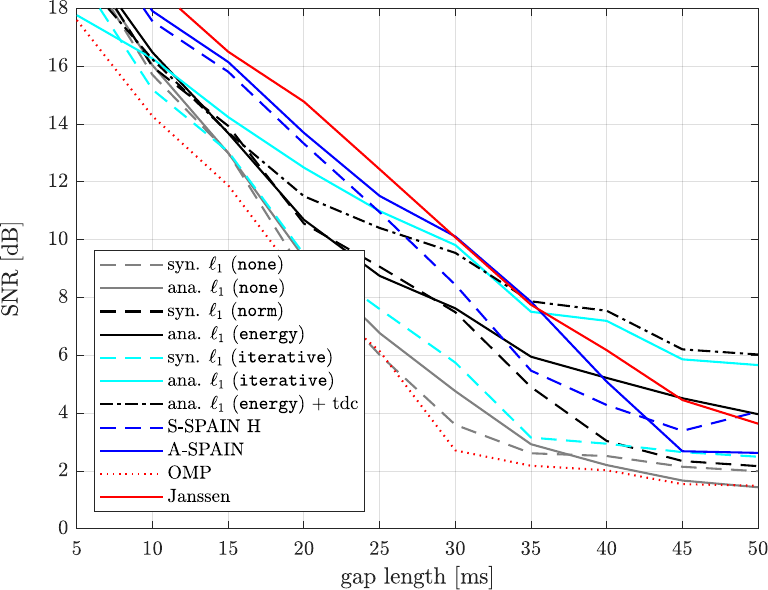}%
			\label{fig:final_test:SNR}}%
	\end{subfloat}%
	\hfill%
	\begin{subfloat}[evaluation by \ODG]{%
			\includegraphics[width=0.47\textwidth]{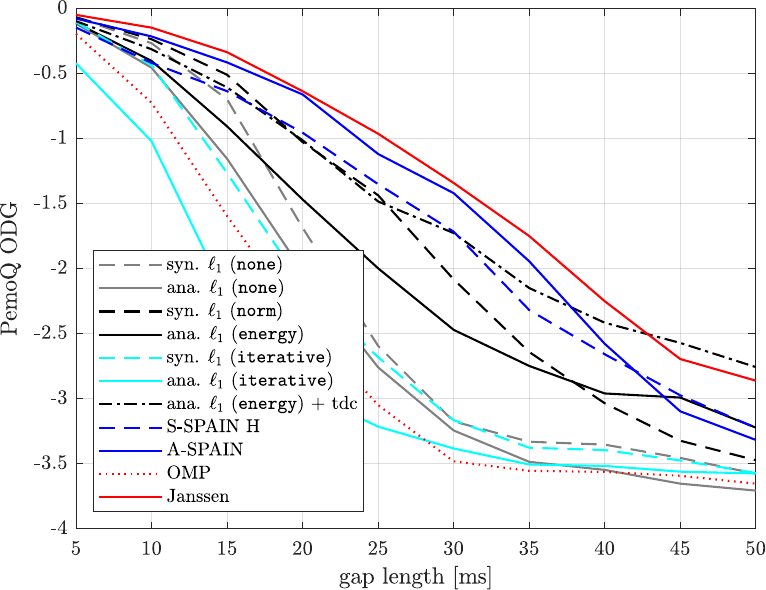}%
			\label{fig:final_test:ODG}}%
	\end{subfloat}%
	\caption{Overall comparison of the methods.}
	\label{fig:final_test}
	\vspace{-0.75em}
\end{figure*}

The evaluation is shown in
Fig.\,\ref{fig:final_test}.
A comparison based on \SNR{} (Fig.\,\ref{fig:final_test:SNR}) reveals %a significant performance
the success of the described analysis model combining weighting the atoms and the time domain processing
(abbreviated as tdc in the figure).
It even outperforms the rather high values of \SNR{} of the iterative reweighting approach.
Such results suggest that the techniques developed are beneficial for the task of compensating the energy drop.
This is apparent for longer gaps in particular,
where the low \SNR{} of the simple $\ell_1$ method
is mainly caused by the disproportion of energy of the original and the restored signal.
For gaps of up to 25\,ms, on the other hand, the non-convex approaches and the Janssen algorithm remained unsurpassed.

Fig.\,\ref{fig:final_test:ODG} shows that the values of \SNR{} in our case mostly coincide with the perceptual measure, the \ODG.
Note, however, the differences regarding the modifications of the $\ell_1$ relaxation.
First, the difference between the analysis and the synthesis model becomes more pronounced in terms of \ODG{} compared to \SNR.
Second, the combined compensation methods do not lead to any major improvement in terms of \ODG{} compared to \SNR.
The interpretation here is that although the time domain compensation
is able to provide energy for the inpainted segment, %the spectrum of the resultant signal is affected in a unnatural way.
the multiplication in the time domain introduces spectral components that are not present in the reference signal.
This effect is then reflected in a~lower \ODG{} value.

Finally, a surprising exception is the iterative reweighting, in which case the values of \ODG{} indicate
an \emph{opposite} result compared to \SNR{} values, especially in the analysis case.
The possible reason is that during the iterations, a~coefficient mistakenly taken as significant in the early phase of the algorithm is amplified by the reweighting procedure in the later phases.
This leads to (audible) artifacts in the restored signal, which is then reflected by the \ODG.
However, testing this hypothesis is beyond the scope of the paper.

\section{Software \& reproducible research}
\label{sec:soft}

\edit{The implementation of the Janssen algorithm was taken from the Audio Inpainting Toolbox % by Adler et\ al.\
\cite{Adler2012:Audio.inpainting}.
OMP was implemented using the Sparsify Toolbox % by Thomas Blumensath
\cite{SparsifyToolbox}.}
%\todo{OMP není ze Smallboxu?}
The MATLAB codes needed for the experiments and all the data and supplemental figures \editt{can be accessed through the web page \url{https://ondrejmokry.github.io/InpaintingRevisited/}}.
\edit{The codes were run in MATLAB versions 2019a and 2020a.}

\section{Conclusion}
We have described the problem of modern optimization-based methods for audio inpainting,
which consists in the lower signal energy in the center of the filled gap.
We have presented a~number of ideas which can effectively deal with this problem
and improve the performance of $\ell_1$-based restoration by bringing more energy to the gap.

The sparse analysis model appears to be more stable in performance with respect to altering the settings of the methods.
%(see the experiment with offset)
Moreover, the analysis model is superior to the synthesis model in most of the cases presented.
Nevertheless, in terms of the SNR, the autoregressive Janssen algorithm can outdo the presented variations in half of the cases, while in terms of ODG, it remains unsurpassed.

In the future, a modification of the Janssen algorithm based on selected ideas from this paper should be considered.

\section*{Acknowledgment}

The work was supported by the joint project of the FWF and the Czech Science Foundation (GA\v{C}R), numbers I\,3067-N30 and 17-33798L, respectively, and GA\v{C}R project number 20-29009S.
Research described in this paper was financed by the National Sustainability Program under grant LO1401.
Infrastructure of the SIX Center was used.

% Can use something like this to put references on a~page
\ifCLASSOPTIONcaptionsoff
  \newpage
\fi

% trigger a~\newpage just before the given reference


% can use a~bibliography generated by BibTeX as a~.bbl file
%\bibliographystyle{IEEEtran}
%\begin{thebibliography}{1}

%\begingroup
%\color{red}
%\endgroup

{
	%\inputencoding{cp1250}
	%\bibliographystyle{IEEEtran}
	%\bibliography{literatura}

	\newcommand{\noopsort}[1]{} \newcommand{\printfirst}[2]{#1}
	\newcommand{\singleletter}[1]{#1} \newcommand{\switchargs}[2]{#2#1}
	
	\vspace{-2em}
	
}

%\end{thebibliography}

% If you have an~EPS/PDF photo (graphicx package needed) extra braces are
% \includegraphics command within an~optional argument. (You could create
% or if you just want to reserve a~space for a~photo:

\begin{IEEEbiography}[{\includegraphics[width=1in,height=1.25in,clip,keepaspectratio]{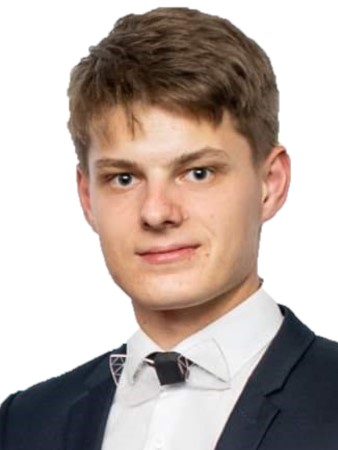}}]{Ondřej Mokrý}
	did his MSc. in Mathematical Engineering at the Faculty of Mechanical Engineering, Brno University of Technology, Czech Republic. Since 2019 he has been pursuing doctoral studies at the Faculty of Electrical Engineering and Communication at the same university. He focuses on applications of sparse regularization, especially on the problem of audio restoration.
\end{IEEEbiography}

\vspace{-2em}

\begin{IEEEbiography}[{\includegraphics[width=1in,height=1.25in,clip,keepaspectratio]{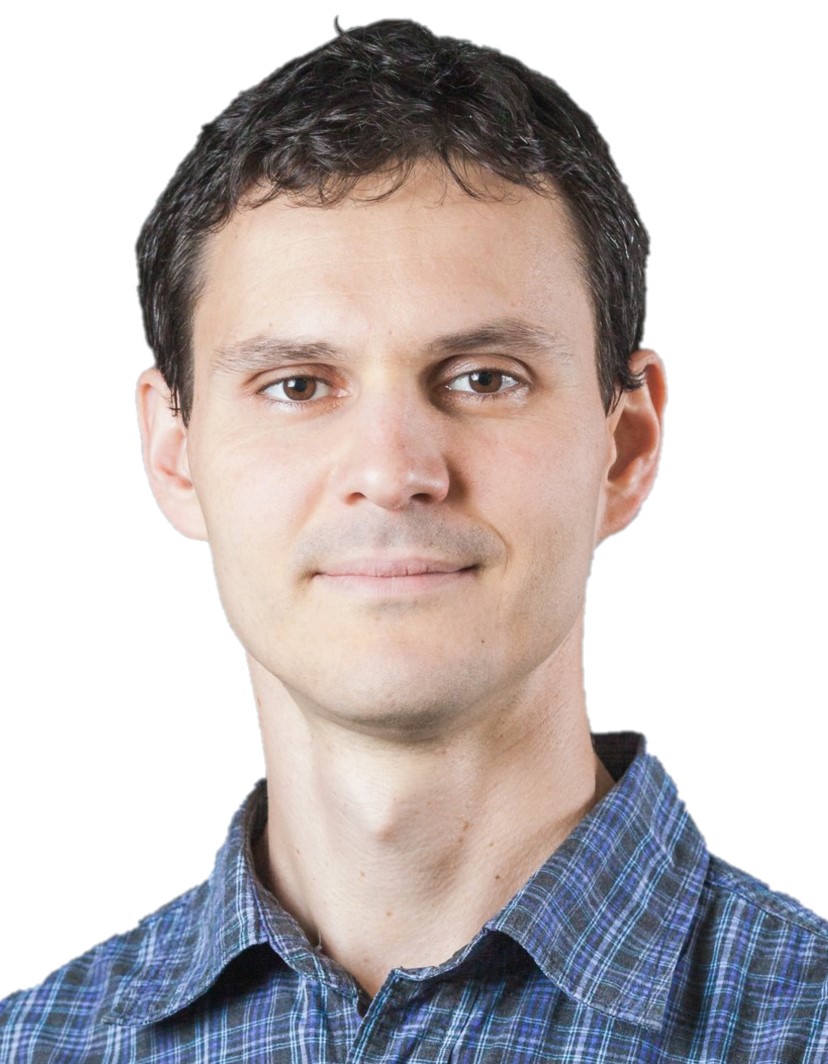}}]{Pavel Rajmic}
	finished his Ph.D. studies in signal processing in 2004 and since then he has been employed at the Faculty of Electrical Engineering and Communication, Brno University of Technology. As a member of the SPLab team, his interests include signal processing, applied and computational mathematics, frame theory and applications, sparse signal modeling, and compressed sensing.
\end{IEEEbiography}

% if you will not have a~photo at all:
%\begin{IEEEbiographynophoto}{Pavel Rajmic}

%\newpage


% a~\vfill before or after them. The appropriate

%\vfill


\end{document}

%% file: defs.tex
\theoremstyle{plain}
\theoremstyle{definition}

\theoremstyle{remark}
\def\RR{\mathbb{R}}
\def\CC{\mathbb{C}}

\def\NN{\mathbb{N}}

\def\x{\vect{x}}
\def\u{\vect{u}}

\def\y{\vect{y}}
\def\z{\vect{z}}
\def\w{\vect{w}}

\def\p{\vect{p}}
\def\q{\vect{q}}

\def\g{\vect{g}}
\def\Mr{M_\mathrm{R}}

\def\grade{g}

\newcommand{\Id}{\mathit{Id}}
\newcommand{\syn}{D}
\newcommand{\ana}{\adjoint{D}}

\newcommand{\norm}[1]{\|#1\|}
\newcommand{\abs}[1]{\left\vert#1\right\vert}

\newcommand{\adjoint}[1]{#1^*}

\newcommand{\vect}[1]{\mathbf{#1}} % vector
\newcommand{\matr}[1]{\mathbf{#1}} % matrix

\newcommand{\ceil}[1]{\left\lceil#1\right\rceil}
\newcommand{\floor}[1]{\left\lfloor#1\right\rfloor}

\newcommand{\argmin}{\mathop{\operatorname{arg~min}}}
\newcommand{\prox}{\mathrm{prox}}
\newcommand{\soft}{\mathrm{soft}}
\newcommand{\clip}{\mathrm{clip}}
\newcommand{\proj}{\mathrm{proj}}

\newcommand{\supp}{\mathrm{supp}}

\newcommand{\offset}{\mathit{offset}}
\newcommand{\SNR}{SNR}

\newcommand{\ODG}{ODG}

\newcommand{\qm}[1]{``#1''} % quotation marks
\newcommand{\edit}[1]{\textcolor{black}{#1}}
\newcommand{\editt}[1]{\textcolor{black}{#1}}